\documentclass[onecolumn,superscriptaddress,
 amsmath,amssymb,
 aps,
 prd,
]{revtex4-1}
\usepackage{dcolumn}
\usepackage{bm}
\usepackage{graphicx}
\usepackage{physics}
\usepackage{comment}
\usepackage{mathtools} 
\usepackage[caption=false]{subfig}
\usepackage{color}
\usepackage{tikz}
\usetikzlibrary{matrix}
\usepackage{pgfplots}

\newcommand{\bea}{\begin{eqnarray}}
\newcommand{\eea}{\end{eqnarray}}

\begin{document}

\title{Resetting Dyson Brownian motion}
\author{Marco Biroli}
\affiliation{LPTMS, CNRS, Univ.  Paris-Sud,  Universit\'e Paris-Saclay,  91405 Orsay,  France}
\author{Satya N. Majumdar}
\affiliation{LPTMS, CNRS, Univ.  Paris-Sud,  Universit\'e Paris-Saclay,  91405 Orsay,  France}
\author{Gr\'egory Schehr}
\affiliation{Sorbonne Universit\'e, Laboratoire de Physique Th\'eorique et Hautes Energies, CNRS UMR 7589, 4 Place Jussieu, 75252 Paris Cedex 05, France}


\begin{abstract}

In this paper, we introduce a new stochastic process of $N$ interacting particles on the line that evolve via Dyson Brownian motion (DBM) with Dyson's index $\beta > 0$ and undergo simultaneous resetting to their initial positions at a constant rate $r$. We call this process the 
{\it resetting Dyson Brownian motion} (RDBM) with a parameter $\beta > 0$, in short the $\beta$-RDBM process. For $\beta = 1,2,4$, the positions of the particles in the RDBM can be interpreted as the eigenvalues of a new random matrix ensemble where the entries of an $N \times N$ Gaussian matrix evolve as simultaneously resetting Brownian motions (with rate $r$) in the presence or absence of a harmonic trap. For $r=0$ and in the presence of a harmonic trap, this system reaches an equilibrium Gibbs-Boltzmann state of the so called Dyson log-gas. However, the stochastic resetting drives the system at long time to a nonequilibrium stationary state (NESS). We compute exactly the joint distribution of the positions of the particles in this NESS for all $\beta>0$ and calculate several macroscopic and microscopic observables in the large $N$ limit. These include the average density profile of the gas, the extreme value statistics, the spacing between two consecutive particles and the full counting statistics, i.e., the distribution of the number of particles in an interval $[-L,L]$. We show that a nonzero resetting rate $r>0$ drastically changes the nature of the fluctuations in the stationary state: while the log-gas ($r=0$) is rather {\it rigid}, the $\beta$-RDBM in its NESS becomes {\it fluffy}, i.e., the fluctuations of different observables are of the same order as their mean. In the absence of a harmonic trap, our results for the $\beta = 2$ RDBM can be related to nonintersecting Brownian motions (vicious walkers) in the presence of resetting. Our model demonstrates interesting effects arising from the interplay between the eigenvalue repulsion and the all-to-all attraction (generated by stochastic resetting) in an interacting particle system. Numerical simulations are in excellent agreement with our analytical results.
 \end{abstract}

\maketitle

\section{Introduction} \label{sec:intro}

Random matrix theory (RMT) is a powerful framework for studying systems with strongly correlated~variables~\cite{W28,W51,M91,F10,LNV18,PB20,MPS20,MSbook}.
%
%
The eigenvalues of random matrices exhibit strong repulsive correlations between each other, the so-called `level spacing'~\cite{W51,M91,F10}. Consider for example an $N \times N$ Gaussian random matrix which can be either real symmetric (like in the Gaussian Orthogonal Ensemble (GOE)), complex Hermitian (the Gaussian Unitary Ensemble (GUE)), or quaternionic self-dual (the Gaussian Symplectic Ensemble (GSE)). These matrices have $N$ real eigenvalues  
which  are statistically equivalent to a gas of $N$ diffusing particles in a harmonic trap with pairwise logarithmic repulsion, the so-called ``Dyson log gas''~\cite{D62,D62b}, which represents a strongly correlated gas of particles in thermal equilibrium. Deriving thermodynamic properties of such a gas is in general very hard, due to the strong correlations between the particles. For example, the extreme value statistics, i.e., the distribution of the largest (or smallest) eigenvalue of such a matrix is highly nontrivial and is not given by any of the usual known Gumbel, Fr\'echet or Weibull distributions. Instead,  
for large $N$, the appropriately centered and scaled distribution of the largest eigenvalue $\lambda_{\max}$ is given by the celebrated Tracy-Widom distribution \cite{TW94,TW96}. Interestingly, this Tracy-Widom distribution has appeared in many different fields~\cite{MS14}. 
These include the longest increasing subsequence of random permutations \cite{BDJ99,BR00,M06}, directed polymers and growth models \cite{DPRM} in the Kardar-Parisi-Zhang (KPZ) universality class in (1 + 1) dimensions as well as for the continuum (1+1)-dimensional KPZ equation \cite{TS10,CLR10,D10,ACQ11}, sequence alignment problems \cite{MN05}, mesoscopic fluctuations in quantum dots \cite{DOTS}, height fluctuations of non-intersecting Brownian motions over a fixed time interval \cite{FMS11,L12}, height fluctuations of non-intersecting interfaces in presence of a long-range interaction induced by a substrate \cite{NM09}, in trapped noninteracting fermions \cite{DLMS15,DLMS16,DLMS19}, and also in finance \cite{BB07}. Remarkably, the Tracy-Widom distributions have been observed in experiments on nematic liquid crystals \cite{KPZ} (for $\beta = 1, 2$), in experiments involving coupled fiber lasers \cite{FPNFD12} (for $\beta  = 1$) and also in disordered superconductors \cite{Lemarie}.

\vspace{0.1cm}
To be precise, let us consider the canonical example of an $N \times N$ random matrix with entries $X_{ij}$ and belonging to the Gaussian ensembles mentioned above. Here, each entry on the diagonal and upper diagonal is sampled randomly from a Gaussian distribution with zero mean and variance chosen such that the joint distribution of the entries can be written in a rotationally invariant form, 
\bea \label{Gauss}
{\rm Prob.}[X] \propto \exp[ - c_0 \Tr(X^\dagger X) ] \;,
\eea
where $c_0$ is a constant that sets the variance of the entries. This probability distribution is invariant under a similarity transformation $X \to S^\dagger X S$ and for such rotationally invariant ensembles, the eigenvalues and eigenvectors decouple and one can just study the joint distribution of the $N$ real eigenvalues $x_1, \cdots, x_N$, without the need to study the associated eigenvectors. In fact, the normalized eigenvectors are just uniformly distributed on the unit $N$-dimensional sphere~\cite{M91}. The joint distribution of the eigenvalues is given by \cite{M91,F10}
\begin{equation} \label{eq:intro-jpdf}
{\rm Prob.}[x_1, \cdots, x_N] \propto \exp[ - c_0 \sum_{i = 1}^N x_i^2 ] \prod_{i < j} |x_i - x_j |^\beta \;,
\end{equation}
where $\beta$ is called the Dyson index with values $\beta = 1, 2$ or 4 corresponding respectively to matrices from the GOE, GUE or GSE ensembles~\cite{M91,F10}. Choosing $c_0 = \beta/2$ the joint law in Eq. (\ref{eq:intro-jpdf}) can be re-written as
\begin{equation} \label{eq:intro-log-jpdf}
{\rm Prob.}[x_1, \cdots, x_N] \propto \exp[ - \beta \left( \frac{1}{2}\sum_{i = 1}^N x_i^2 -  \frac{1}{2}\sum_{i \neq j} \ln |x_i - x_j| \right) ] \propto e^{-\beta E[\{x_i\} ]} \;.
\end{equation}
Even though this result for the joint distribution of eigenvalues came originally from rotationally invariant ensembles with $\beta = 1,2, 4$, it turns out that the same joint distribution for arbitrary $\beta > 0$ also arises for a special class of tri-diagonal matrices, with off-diagonal entries distributed as chi-square variables and parametrized by $\beta$ \cite{DE02}. This joint law can be interpreted as the Gibbs-Boltzmann measure of an interacting gas of logarithmically repellent particles on a line subjected to a confining parabolic potential with total energy $E[\{x_i\} ]$. The Dyson index $\beta$ plays the role of the inverse temperature: this is known as the {\it Dyson log-gas} \cite{D62,D62b}. Given this explicit joint distribution, the challenge in RMT is to compute several macroscopic and microscopic observables, such as the average density of eigenvalues, the spacing distribution between consecutive eigenvalues, the extreme value statistics (EVS), i.e., 
the distribution of the largest eigenvalue, the full counting statistics, i.e., the distribution of the number of eigenvalues in a given interval. These computations are in general quite difficult for finite $N$, except for $\beta = 2$ (GUE), which has a determinantal structure that allows for exact calculations for any finite~$N$ \cite{M91,F10}. However, for large $N$, one can make progress for general $\beta$ using Coulomb gas techniques and many results are well known in the literature \cite{M91,F10}. Apart from this log-gas, there are very few examples of solvable strongly correlated gases in confinement, an example being the Riesz gas in one-dimension where the pairwise interaction behaves as a power law of the distance between two particles~\cite{Agar19,Lew22,Ket24}. 

Another way to arrive at this joint distribution in Eq.~(\ref{eq:intro-jpdf}) for $\beta = 1,2, 4$ was originally due to Dyson \cite{D62}. He considered each entry of the matrix $X_{j,k}(t)$ performing an independent Ornstein-Uhlenbeck (OU) process in a harmonic potential of stiffness $\mu$ as a function of a fictitious time $t$. For $\mu=0$, each entry then performs an independent Brownian motion, instead of an OU process. We will generally refer to this stochastic process with $\mu >0$ as a {\it matrix OU process}.
In the long time limit $t \to \infty$ and $\mu >0$, the matrix entries reach a stationary state given by Eq. (\ref{Gauss}) where $c_0$ is related to the stiffness and the diffusion constant of the OU process. One can also study how the $N$ real eigenvalues evolve as a function of this fictitious time $t$. In fact, one can write down effective coupled  Langevin equations for the eigenvalues (see later in Eq. (\ref{Langevin})), whose stationary state is then given by Eq. (\ref{Gauss}). This is the equilibrium measure of the Dyson log-gas, where the repulsion between any pair of eigenvalues is manifest via the logarithmic interaction term. The evolution of the eigenvalues in this fictitious time $t$ is usually referred to as {\it Dyson Brownian motion} (DBM), which has been studied in many different applications~\cite{KT04,Bor09,ABG12,GMS21,MM21,TLS23}. Although these Langevin equations were originally derived as the time-evolution of eigenvalues for Gaussian random matrices with $\beta = 1,2,4$, they can be studied for arbitrary $\beta>0$ as a general interacting particle system without the interpretation of being eigenvalues of an underlying Gaussian random matrix. This interacting particles system is usually referred to as the $\beta$-DBM.

In a completely different context, a new exactly solvable model with strong long-range correlations was recently introduced based on stochastic resetting \cite{BLMS23}. Stochastic resetting simply means interrupting the natural dynamics of a system at random times and instantaneously restarting the process either from its initial configuration or more generally from any pre-decided state. The interval between two successive resettings is typically Poissonian, although other protocols such as periodic resetting have also been studied. One of the main effects of stochastic resetting is that the resetting moves violate detailed balance and drive the system to
a non-equilibrium stationary state (NESS) \cite{EM11, EM11b} -- for recent reviews see \cite{EM20,PKR22,GJ22}. Initially, stochastic resetting was studied mostly in the context of a single particle with different dynamics, but subsequently has been extended to systems with many degrees of freedom \cite{GMS14,DHP14,BKP19,MMS20,NG23}. In Ref. \cite{BLMS23}, a system of $N$ non-interacting Brownian motions on the line was subjected to simultaneous resetting. This means that each of the Brownian walkers evolves independently up to a random time $\tau$ distributed via a $p(\tau) = r\, e^{-r\tau}$ and then they are reset simultaneously back to the origin and the process renews itself after this. It was shown that this simultaneous resetting leads to the dynamical emergence of strong attractive correlations between particles, even though there are no direct interactions between them. These strong correlations persist all the way up to the stationary state and the joint distribution of the positions of the particles in the stationary state can be computed exactly \cite{BLMS23}. This stationary joint distribution is not factorizable into individual marginal distributions, reflecting the correlations between the particles. 
Despite the fact that the particles are strongly correlated, the stationary state still retains an exactly solvable structure and several observables of this gas (both macroscopic as well as microscopic as mentioned earlier) can be computed exactly. There are two important differences between this resetting gas and the standard Dyson log gas: (i) the effective interaction in the stationary state between the particles in the resetting gas is all-to-all, as opposed to ``pairwise'' in the log-gas and (ii) the effective interaction is attractive in the resetting gas, while they are repulsive in the log-gas. This mechanism for generating strong {\bf attractive} correlations via simultaneous resetting can be generalized to a wider class of systems, both classical and quantum, 
that all share a common structure of the joint distribution of particle positions in the stationary state where they are conditionally independent and identically distributed, the so called (CIID) variables \cite{BLMS24,BKMS24,SM24, KMS24}.


This brings us to the main topic of this paper, namely what happens if the $\beta$-DBM process is subject to simultaneous Poissonian resetting with rate $r$. In other words, we consider a system of $N$ interacting particles that evolve via the $\beta$-DBM dynamics up to a random time $\tau$ distributed via $p(\tau) = r\,e^{-r \tau}$ and then the particles are simultaneously reset to their initial positions and the process restarts. We call this resetting process the $\beta$-RDBM and  
our goal is to investigate the joint distribution ${\rm Prob.}[x_1, \cdots, x_N | r]$ of the positions of the particles in the stationary state of the $\beta$-RDBM process in the long time limit $t \to \infty$. We recall that, in the absence of resetting $r=0$, the particles {\it repel} each other via pairwise logarithmic interactions, as in Eq. (\ref{eq:intro-log-jpdf}). On the other hand, when simultaneous resetting with rate $r>0$ is switched on, it induces an all-to-all {\it attraction} between the particles. For the special values $\beta = 1,2,4$, we can interpret the $\beta$-RDBM process as the stochastic evolution of the eigenvalues of a $N \times N$ Gaussian random matrix, whose entries perform independent Ornstein-Uhlenbeck processes with stochastic resetting. In this $\beta$-RDBM model, there is a competition between the pairwise repulsion and the all-to-all attraction and it is interesting to study how the behavior of different observables in the stationary state get affected by this competing {\bf repulsive and effectively attractive} interactions. Motivated by the studies of eigenvalues in RMT, we will study in this paper the following observables: 
\begin{itemize}
\item The average particle density in the NESS
\begin{equation} \label{eq:intro-def-rho}
\rho_N(x | r) = \left\langle \frac{1}{N}\sum_{i = 1}^N \delta(x - x_i) \right\rangle \;,
\end{equation}
where $\langle \cdot \rangle$ denotes an average over the stationary joint distribution ${\rm Prob.}[x_1, \cdots, x_N | r]$. Note that, by definition, the average density is normalized to unity: $\int_{-\infty}^\infty  \rho_N(x | r)\, dx = 1$. Due to the presence of the confining harmonic potential, the gas is not translationally invariant. On an average, one would expect a higher density of particles near the trap center and the density should decrease as one goes away from the trap center. The quantity $\rho_N(x | r)$ in Eq. (\ref{eq:intro-def-rho}) thus provides the average density profile of the gas. For example, for $r=0$ and $\beta >0$, the appropriately scaled density profile is given, for large $N$, by the celebrated Wigner semi-circular law \cite{M91}. We will see here that, for nonzero $r>0$, the Wigner semi-circular law gets completely modified and we will provide an exact expression of this average density in the large $N$ limit for all $\beta >0$. For a schematic representation of the average density profile of the gas for $r=0$ (standard Dyson log-gas) and $r>0$ ($\beta$-RDBM), see Fig. \ref{fig:sketch}.

\item We will also compute exactly the distribution of the position of the rightmost particle in the NESS, i.e., for the random variable  
$x_{\rm max} = \max \{x_1, x_2, \cdots, x_N\}$. For $r=0$ and $\beta >0$, the centered and scaled distribution of $x_{\max}$, for large $N$, is given
by the Tracy-Widom distribution with parameter $\beta$ \cite{TW94,TW96}. For $r>0$, we will see that the scaled distribution of $x_{\max}$ for large $N$ gets completely modified and is given by a new distribution, very different from the Tracy-Widom law (see, e.g., Fig. \ref{fig:sketch}).

\item Next we will study the distribution of the spacing between two consecutive particles in the NESS. For $r=0$ and $\beta = 1,2,4$, i.e., for the standard Gaussian RMT ensembles, the scaled spacing distribution is known for large $N$ in terms of complicated Painlev\'e functions \cite{F10}. However, Wigner found that these limiting distributions are very well approximated by the distribution of the scaled spacing between two eigenvalues of a $2 \times 2$ Gaussian random matrix, which is much easier to compute. This is famously known as Wigner's surmise \cite{Surmise}. Following this ansatz, we will study the scaled spacing distribution for $r>0$ and $\beta>0$ for a $2$-particle $\beta$-RDBM. We will show that, indeed, the Wigner's surmise works very well also for $r>0$ by comparing our analytical results with numerical simulations for large $N$.

\item The next observable in our list is the full counting statistics (FCS), defined as the distribution of the number of particles $N_L$ in an interval $[-L,L]$ 
in the NESS, as marked schematically in Fig. \ref{fig:sketch}. Once again, for $r=0$ and $\beta >0$, the result for the FCS for large $N$ is well known in the RMT literature~\cite{DM63,CL95,FS95,MMSV14,CLM15,MMSV16}. Here we will compute it exactly for $r>0$ and $\beta > 0$ and show this is drastically different from that of the RMT case (i.e., $r=0$). 

\end{itemize}

We will further show an interesting by-product of our results. Our analytical results are valid for an arbitrary stiffness $\mu$ of the confining harmonic potential of the log-gas. For the special case $\beta = 2$, by setting $\mu = 0$ (i.e., un-trapping the gas), we will show that our results can be used to describe $N$ non-intersecting Brownian motions on a line \cite{M1984,SMCR08,Bor09,RS11,GMS21}, but with the new twist of having simultaneous resetting close to the origin with rate $r$.

\vspace{0.1cm}
The rest of the paper is organised as follows. In Section \ref{sec:intro-model} we introduce our model, i.e., the $\beta$-RDBM process
and provide a brief summary of the main results in Section \ref{sec:intro-main-results}. Section \ref{sec:model} provides the detailed derivations for the four observables: the average density profile (\ref{sec:model-density}), the EVS (\ref{sec:model-evs}), the spacing distribution (\ref{sec:model-gap}) and the FCS (\ref{sec:model-fcs}). Finally we conclude in Section~\ref{sec:conclusion}. In Appendix \ref{App} we derive the exact propagator of the $\beta$-DBM process at all times.

\section{The model and the main results}

\subsection{The model} \label{sec:intro-model}

\begin{figure}
\centering
\includegraphics[width = 0.45\textwidth]{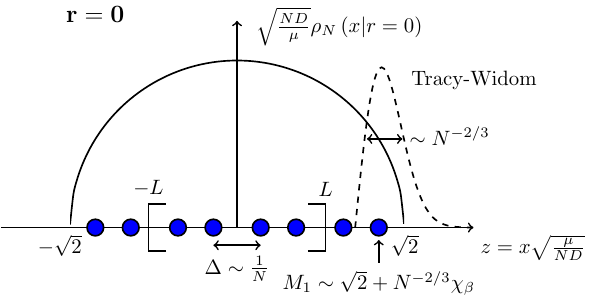}\hspace*{1.5cm}\includegraphics[width = 0.45\textwidth]{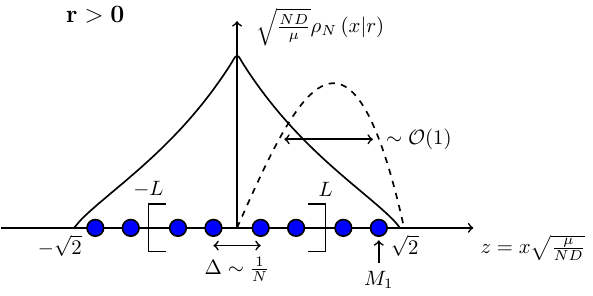}
\caption{A schematic representation of the positions of the particles in the log-gas, i.e., $r=0$ (left panel) and the RDBM in its NESS, i.e., $r>0$ (right panel). 
The scaled average density profile, in both cases, is supported over $z \in [-\sqrt{2},+\sqrt{2}]$, but the shape is rather different: 
on the left it has a semi-circular shape, while on the right it has a cusp at $z=0$, where the simultaneous resetting takes place. The distribution of the 
position $x_{\max}$ of the rightmost particle is shown schematically by a dashed curve in both figures and they have very different behaviors. On the left, the distribution is peaked around $\sqrt{2}$ with a width of order $O(N^{-2/3})$ (the Tracy-Widom distribution), while, on the right, the distribution is supported over $[0,\sqrt{2}]$. The inter-particle spacing $\Delta$ is of order $O(1/N)$ in scaled units in both models.} \label{fig:sketch}
\end{figure}

Here we first recall the standard log-gas and the associated $\beta$-DBM in the absence of resetting. Then we will define our $\beta$-RDBM process 
by introducing simultaneous resetting with rate $r$. 

\vspace*{0.5cm}
\noindent {\bf Log-gas and DBM without resetting $r=0$.} We start by recalling some known results from RMT and the Dyson log-gas \cite{D62}. 
For simplicity we introduce the DBM for the GOE (with Dyson's index $\beta = 1$) and later generalise it to $\beta=2$ and $\beta = 4$. Let us consider a real symmetric $N \times N$ matrix whose entries $X_{j,k}$'s perform independent Orhnstein-Uhlenbeck (OU) processes in some fictitious time $t$, i.e.
{
\begin{equation} \label{eq:dynamics-GUE}
\dv{X_{j, k}(t)}{t} = - \mu X_{j, k}(t) + \sqrt{(1 + \delta_{jk}) D } \, \eta_{jk}(t)
\end{equation}
}
where 
{
\begin{equation}\label{eq:eta-mu-def}
\langle \eta_{jk}(t) \eta_{mn}(t') \rangle = \delta_{jm} \delta_{kn} \delta(t - t') \;.
\end{equation}
We assume that the process starts at the origin $X_{j,k}(t=0)=0$. Thus $X_{j,k}$ can be thought of as the position of a particle labelled by $(j,k)$ on the real line, performing Brownian motion in the presence of a confining potential $\mu X_{j,k}^2/2$. The position distribution at any time $t$ is a simple Gaussian
\begin{equation} \label{eq:OU-propagator-sigma}
{\rm Prob.}[X_{j,k}(t)] = \frac{1}{\sqrt{\pi \sigma^2(t) (1 + \delta_{jk})}} \exp( - \frac{X_{j,k}(t)^2}{\sigma^2(t) (1 + \delta_{jk})} ) \;,
\end{equation} 
where 
\begin{equation} \label{eq:def-sigma}
\sigma^2(t) = \frac{D (1 - e^{-2 \mu t})}{\mu} \;.
\end{equation}
The distribution of the full matrix is obtained by taking the product of the distributions of all its independent components as given in Eq. (\ref{eq:OU-propagator-sigma}), which yields
\begin{equation} \label{eq:matrix-propagator}
{\rm Prob.}[X(t) = X] = \left( \frac{1}{2 \pi \sigma^2(t)} \right)^{N/2} \left(\frac{1}{\pi \sigma^2(t)} \right)^{N(N-1)/4} \exp( - \frac{\Tr[X^2]}{2 \sigma^2(t)} ) \;.
\end{equation}
The prefactors take the form above because the degrees of freedom of the matrix are $N$ diagonal elements of variance $\sigma^2(t)$ and $N(N-1)/2$ off-diagonal elements of variance $\sigma^2(t)/2$. In the long-time limit, the variance $\sigma^2(t)$ reaches a stationary value $\sigma^2(t \to +\infty) = \frac{D}{\mu}$ and hence the process in Eq. (\ref{eq:matrix-propagator}) reaches an equilibrium steady state
\begin{equation} \label{eq:matrix-ss}
{\rm Prob.}[X(t \to \infty) = X] = \frac{1}{2^{N/2}} \left( \frac{\mu}{\pi D}\right)^{\frac{N(N+1)}{4}} \exp( - \frac{\mu}{2D} \Tr[X^2]  ) \;.
\end{equation}
}
The $N$ real eigenvalues associated to this Gaussian matrix have the joint distribution~\cite{M91,F10} 
\bea \label{P_joint_Xeq_GOE}
{\rm Prob.}[x_1, \cdots x_N] \propto e^{- \frac{\mu}{2D} \sum_{i=1}^N x_i^2} \prod_{i<j} |x_i - x_j| \;,
\eea
where the proportionality constant can be fixed from the normalisation condition $\int dx_1 \dots \int dx_N {\rm Prob.}[x_1, \cdots x_N] =1$. 

\vspace*{0.3cm}
Following exactly the same method as for $\beta = 1$, one can similarly compute the distribution of the matrix $X(t)$ at any time $t$ for $\beta= 2$ and $\beta = 4$ where the entries are respectively complex and quaternions. Skipping details, one finds that for any $\beta=  1, 2, 4$, the distribution of the matrix remains Gaussian at any time $t$ and is given by
\begin{equation} \label{eq:matrix-propagator_beta}
{\rm Prob.}[X(t) = X] \propto \exp( - \beta \, \frac{\Tr[X^\dagger X]}{2 \sigma^2(t)} ) \;,
\end{equation}
where we have introduced the factor $\beta$ inside the exponential for later convenience. Taking the limit $t \to \infty$ in Eq. (\ref{eq:matrix-propagator_beta}), the stationary distribution of the matrix for $\beta = 1, 2, 4$ reads
\bea
{\rm Prob.}[X(t \to \infty) = X] \propto \exp( - \frac{\beta \mu}{2D} \Tr[X^\dagger X]  ) 
\eea
and the associated joint distribution of eigenvalues is given by \cite{F10}
\bea \label{P_joint_Xeq_Gbeta}
{\rm Prob.} [x_1, \cdots, x_N] \propto e^{- \frac{\beta \mu}{2D} \sum_{i=1}^N x_i^2} \prod_{i<j} |x_i - x_j|^\beta \propto e^{-\beta E[\{ x_i\}]} \;.
\eea
Here the energy $E[\{ x_i\}]$ associated with 
the positions $\{x_i\}$'s of the gas is given by
\bea \label{energy}
E[\{ x_i\}] = \frac{\mu}{2D} \sum_{i=1}^N x_i^2 - \frac{1}{2} \sum_{i \neq j} \ln |x_i-x_j| \;.
\eea
Thus, in the stationary state, the eigenvalues form a gas of $N$ particles which is at thermal equilibrium with $\beta$ playing the role of the
inverse temperature. The energy of the gas in Eq. (\ref{energy}) has two components: the first term corresponds to an external harmonic
potential $\mu/(2 D) x_i^2$, while the second term represents pairwise logarithmic repulsion between particles. Hence this gas is referred to
as the ``Dyson log gas''.

\vspace*{0.3cm}
An alternative way to arrive at the same stationary state of the eigenvalues, for $\beta = 1,2,4$, is to consider an overdamped Langevin equation~\cite{D62} 
\begin{equation} \label{Langevin}
\dv{x_i}{t} = - \mu x_i + D \sum_{j (\neq i)} \frac{1}{x_i - x_j} + \sqrt{\frac{2 D}{\beta}} \, \eta_i(t) \;,
\end{equation}
where $\eta_i(t)$ is a Gaussian white noise with $\langle \eta_i(t) \rangle = 0$ and $\langle \eta_i(t) \eta_j(t') \rangle = \delta_{ij} \delta(t - t')$. We assume that the Langevin equations in Eq. (\ref{Langevin}) start from an initial condition where all the particles are localised close to the origin, e.g., over an interval $[-\epsilon,+\epsilon]$ with 
\bea \label{IC}
x_i(0) = \epsilon\left(-1 + \frac{2i}{N} \right) \;, \; i =1,2, \cdots, N \;.
\eea
Eventually, we will take the limit $\epsilon \to 0$. Thus the eigenvalues can be interpreted as the positions of $N$ particles on a line, each performing OU process in the presence of a pairwise repulsive force. Let ${\rm Prob.}[x_1, x_2, \cdots, x_N,t] \equiv p(\vec{x}),t$ denote the joint distribution of the positions of the particles at time $t$, i.e., the propagator at time $t$. Its time evolution is governed by the Fokker-Planck equation associated to the Langevin equation (\ref{Langevin}) and it reads
\begin{equation} \label{eq:FP-DBM_text}
\pdv{p(\vec{x}, t)}{t} = \frac{D}{\beta} \sum_{i = 1}^N \pdv[2]{p(\vec{x}, t)}{x_i} - \sum_{i = 1}^N \pdv{}{x_i} \left[\left(-\mu x_i + D \sum_{j \neq i} \frac{1}{x_i - x_j} \right) p(\vec{x}, t) \right] \;.
\end{equation}
By setting $\partial{p(\vec{x}, t)}/{\partial t} = 0$ as $t \to \infty$, it is easy to check that the stationary joint distribution ${\rm Prob.}[x_1, x_2, \cdots, x_N] = \lim_{t \to \infty} {\rm Prob.}[x_1, x_2, \cdots, x_N,t]$ is indeed given by the Gibbs state in Eq. (\ref{P_joint_Xeq_Gbeta}). Note that for $\beta = 1,2,4$ the positions $x_i$'s can be interpreted as the eigenvalues of an underlying Gaussian matrix. However, the Langevin equation (\ref{Langevin}) is well defined for any $\beta > 0$. But for $\beta \neq 1,2,4$, they do not have any interpretation as the eigenvalues of an underlying Gaussian matrix. We refer to this Langevin process in Eq. (\ref{Langevin}) as the~$\beta$-DBM.

\vspace*{0.5cm}
\noindent {\bf DBM with simultaneous resetting with rate $r>0$.} We can introduce stochastic resetting in DBM in two alternative ways: 
\begin{itemize}

\item[(i)] For $\beta=1,2,4$, we can start from an $N \times N$ Gaussian matrix whose elements perform independent OU processes, as in Eq. (\ref{eq:dynamics-GUE}), starting from $X_{j,k}(t=0)=0$, but with a constant rate $r$, they are {\it simultaneously} reset to the origin. This is equivalent to evolving the matrix entries via Eq. (\ref{eq:dynamics-GUE}) up to a certain random time $\tau$ drawn from an exponential distribution $p(\tau) = r\, e^{- r \tau}$ and then resetting them simultaneously to the initial condition close to the origin as in Eq.~(\ref{IC}), followed by a restart of the dynamics. In other words, in a small time interval $\dd t$, with probability $1 - r \dd t$ the entries evolve via independent OU processes, while with the complementary probability $r \dd t$ one resets all matrix entries to their initial value $0$ simultaneously. At any instant of time, one can diagonalise this matrix and observe the evolution of the eigenvalues. Between any two consecutive resettings, the eigenvalues will evolve via the DBM in Eq. (\ref{Langevin}) and then get reset to their initial conditions. We will refer to this process of eigenvalues as a resetting Dyson Brownian motion (RDBM) with $\beta = 1,2,4$.

\item[(ii)] For general $\beta > 0$, we can think of a set of particles with positions $x_i$'s that evolve via the $\beta$-DBM in Eq. (\ref{Langevin}) and then get simultaneously reset to their initial conditions with rate $r$. For a general $\beta>0$, the $x_i$'s do not have the interpretation of the eigenvalues but, nevertheless, they represent a system of harmonically confined particles in one-dimension with pairwise repulsion and stochastic resetting. This is a wider class of models parametrized by $\beta > 0$ and for the special cases $\beta = 1,2,4$ it has the interpretation of evolving eigenvalues with stochastic resetting. 
For general $\beta$, we will refer to this process as the $\beta$-RDBM. 

\end{itemize}

In this paper, we will focus on the $\beta$-RDBM process. Our principal goal is to derive the joint distribution of the positions of the particles in the stationary state of this $\beta$-RDBM process. Let us define this joint distribution at time $t$ by ${\rm Prob.}[x_1,\cdots, x_N,t \vert r]$, where $r$ denotes the resetting rate. Using a renewal approach, it is possible to express this joint distribution in terms of the propagator ${\rm Prob.}[x_1,\cdots, x_N,t \vert r=0]$ of the reset-free $\beta$-DBM evolving via Eq. (\ref{Langevin}). In fact, we can write down an exact renewal equation
\bea \label{renew_x_t}
{\rm Prob.}[x_1,\cdots, x_N,t \vert r] = e^{-rt}\, {\rm Prob.}[x_1,\cdots, x_N, t \vert r = 0] + r \int_0^t \dd \tau \, e^{-r \tau}\, {\rm Prob.}[x_1,\cdots, x_N, \tau \vert r = 0] \;, 
\eea
This renewal form can be understood as follows. In a fixed time $t$, there are either no resetting or one or more resetting events. The former event (no resetting) gives rise to the first term on the right hand side (RHS) of Eq. (\ref{renew_x_t}). Indeed, the probability that there are no resettings in the interval $[0,t]$ is simply $e^{-r t}$ and in this case the position distribution is trivially the one in the absence of resetting. This explains the first term. The second term on the RHS of Eq. (\ref{renew_x_t}) corresponds to when one or more resetting events occur in $[0,t]$. In this case, it is convenient to consider the interval $\tau$ elapsed since the last resetting before time $t$, since we do not care of what happened before the last resetting. During this interval $\tau$, which itself is random, the entries evolve up to a duration $\tau$ without resetting, i.e. $r=0$, starting from the origin, explaining the factor ${\rm Prob.}[x_1,\cdots, x_N, \tau \vert r = 0] $ inside the integral on the RHS of Eq. (\ref{renew_x_t}). The probability that this interval has duration in $[\tau, \tau + \dd\tau]$ is simply $r\,e^{-r\tau} \,\dd \tau$. Finally, integrating over all $\tau \in [0,t]$ one gets the second term on the RHS of Eq. (\ref{renew_x_t}). 
In the long time limit $t \to \infty$, the first term on the RHS of Eq.~(\ref{renew_x_t}) drops out and one gets the stationary joint distribution of the positions   
\bea \label{renew_x}
{\rm Prob.}[x_1,\cdots, x_N \vert r] =\lim_{t \to \infty}{\rm Prob.}[x_1,\cdots, x_N,t \vert r]= r \int_0^\infty \dd \tau \, e^{-r \tau}\, {\rm Prob.}[x_1,\cdots, x_N, \tau \vert r = 0] \;. 
\eea
Thus to compute the joint distribution of the $\beta$-RDBM in the NESS we need the propagator ${\rm Prob.}[x_1,\cdots, x_N, \tau \vert r = 0]$ of the reset-free $\beta$-DBM {\it at all times $\tau$}, and not just at late times. Hence we need the solution of the Fokker-Planck equation (\ref{eq:FP-DBM_text}) at all times $\tau$. Interestingly, this exact solution for all $\tau$ and all $\beta >0$ can be explicitly written down as
\bea \label{propag_DBM_intro}
{\rm Prob.}[x_1,\cdots, x_N, \tau \vert r = 0]= \frac{1}{Z_N(\beta)}
\frac{1}{\sigma(\tau)^{N + \beta N(N-1)/2}} \exp[ -\frac{\beta}{2}\left(\frac{1}{\sigma^2(\tau)} \sum_{i = 1}^N x_i^2 - \sum_{i \neq j} \ln |x_i - x_j| \right)] \;,
\eea
where $\sigma^2(\tau) = {D (1 - e^{-2 \mu \tau})}/{\mu}$ is already defined in Eq. (\ref{eq:def-sigma}) and $Z_N(\beta)$ is a normalisation constant that can be computed explicitly. Note that while for $\beta = 1,2,4$ this exact propagator of the DBM at any time $\tau$ was well known by exploiting its connection to the rotationally invariant Gaussian matrices (see e.g. \cite{F10}), for general $\beta$ we have not seen this result in the literature. Hence in Appendix \ref{App}, we provide the derivation of this expression in Eq. (\ref{propag_DBM_intro}) for arbitrary $\beta$. Substituting Eq. (\ref{propag_DBM_intro}) into Eq. (\ref{renew_x}), we get the exact joint distribution of the $\beta$-RDBM process in its NESS
\begin{equation} \label{eq:eigenvalues-jpdf}
{\rm Prob.}[x_1, \cdots, x_N | r ] = \frac{1}{Z_N(\beta)} \int_0^{+\infty} r \dd\tau\; e^{-r\tau} \frac{1}{\sigma(\tau)^{N + \beta N(N-1)/2}} \exp[ -\frac{\beta}{2}\left(\frac{1}{\sigma^2(\tau)} \sum_{i = 1}^N x_i^2 - \sum_{i \neq j} \ln |x_i - x_j| \right)] \;.
\end{equation}
We can rewrite this joint distribution as 
\begin{equation}\label{eq:eigenvalues-jpdf-E}
{\rm Prob.}[x_1, \cdots, x_N | r] = \frac{1}{Z_N(\beta)} \int_0^{+\infty} r \dd \tau \;  \frac{e^{-r \tau}}{\sigma(\tau)^{N + \beta N(N-1)/2}} \exp[ - \beta E\left[ x_1, \cdots, x_N; \tau \right] ] \;,
\end{equation}
where $E(x_1, \cdots, x_N; \tau)$ can be interpreted as the energy of a Coulomb gas parametrized by $\tau$ 
{
\begin{equation} \label{eq:def-Z-E}
E[x_1, \cdots, x_N; \tau] = \frac{1}{2 \sigma^2(\tau)} \sum_{i = 1}^N x_i^2 - \frac{1}{2} \sum_{i \neq j} \ln |x_i - x_j| \;.
\end{equation}
Here the $\tau$ dependence arises from the factor $\sigma^2(\tau) = D(1-e^{-2 \mu \tau})/\mu$. Eqs. (\ref{eq:eigenvalues-jpdf-E}) and (\ref{eq:def-Z-E}) form the starting point of our analysis. Note that in the limit $r \to 0$, we can rescale $r \tau = u$ in Eq. (\ref{eq:eigenvalues-jpdf}) and one recovers the standard log-gas with the joint distribution of eigenvalues 
\bea \label{GOE}
{\rm Prob.}[x_1, x_2, \cdots, x_N \vert r= 0] = \frac{1}{Z_N(\beta)} \left(\frac{\mu}{D}\right)^{N + \beta\frac{N(N+1)}{2}} e^{-\beta E[\{x_i\}]} \;,
\eea 
where the energy $E[\{x_i\}]$ of the log-gas is given by Eq. (\ref{energy}).

\vspace*{0.3cm}
Note again that for the three special values $\beta = 1,2, 4$ the stationary distribution of the positions of the particles can be interpreted as the eigenvalues of a rotationally invariant $N \times N$ Gaussian matrix $X$ whose entries are distributed via
\begin{align} 
{\rm Prob.}&[X   | r] \propto \int_0^{+\infty} r \dd \tau \; \frac{e^{- r \tau} }{\sigma_\beta(\tau)^{N + \beta N(N-1)/2}} \exp( - \beta\,\frac{\Tr[ X^\dagger X ]}{2 \sigma^2(\tau)}) \;.\label{eq:eigenvalues-jpdf-beta}
\end{align}
where $\sigma^2(\tau)  = D (1 - e^{-2 \mu \tau})/\mu$, as defined in Eq. (\ref{eq:def-sigma}). This probability measure does not factorise into independent Gaussian for each entry. This is due to the simultaneous resetting of all the values which induces strong correlations between all the components. However, Eq. (\ref{eq:eigenvalues-jpdf-beta}) shows that the distribution is still spherically symmetric, i.e. rotationally invariant, since it is only a function of the trace of the matrix. 
Such deformed Gaussian random matrix ensembles have appeared before in the literature under the name ``super-statistics'' where the variance of the matrix entries is considered as a random variable with some distribution \cite{AM05,BCP08,AAV09}. However, in these papers there was no ``microscopic'' dynamics involved that led to such deformed ensembles and the distribution of the variance was put in by hand. In contrast, the deformed Gaussian ensemble in Eq. (\ref{eq:eigenvalues-jpdf-beta}) for the three special values $\beta = 1,2,4$ appears naturally via the underlying stochastic resetting. Moreover, our $\beta$-RDBM model for the evolution of $x_i$'s is more general and valid for arbitrary $\beta >0$ and not just restricted to $\beta = 1,2,4$.

\vspace*{0.3cm}   
\noindent{\bf Connection to vicious Brownian motions with simultaneous resetting for $\beta = 2$}. One can show that in the special $\beta =2$ and $\mu=0$ (un-trapped particles) of our RDBM model, there is a close connection with another model, namely vicious Brownian motions (VBM) with simultaneous resetting. VBM refers to a system of $N$ Brownian motions conditioned not to intersect each other (the word ``vicious'' refers to the fact that if any pair of particles intersect each other, they kill each other). Consider first the case $r=0$, $\mu = 0$ and $\beta = 2$. In this case, it is well known that the propagators of the DBM and the VBM are related to each other via  
(see e.g. \cite{GMS21,RS11})
\begin{equation} \label{eq:dbm-vicious-link}
{\rm Prob.}^{\rm DBM}_{\beta = 2, \mu=0}[\bm{x}, t | \bm{x_0}, 0] = \frac{\prod_{i < j} (x_i - x_j) }{\prod_{i < j} (x_{0i} - x_{0j}) } {\rm Prob.}^{\rm VBM}[\bm{x}, t | \bm{x_0}, 0] \;,
\end{equation}
{where ${\rm Prob.}_{\beta = 2, \mu=0}^{\rm DBM}[\bm{x}, t | \bm{x_0}, t_0]$ denotes the probability for a DBM with $\beta = 2$ and $\mu=0$, 
starting at $\bm{x_0}$ at time $t_0$, to reach $\bm{x}$ at time $t$ (and equivalently for the VBM).}
In the presence of resetting Eq. (\ref{renew_x_t}) allows us to express the RDBM propagator in the NESS as 
\begin{equation} \label{eq:resetting-free-link}
{\rm Prob.}^{\rm RDBM}[\bm{x} | r, \bm{x}_{\rm reset} = \bm{0}] = r \int_0^{+\infty} \dd \tau \; e^{-r \tau} {\rm Prob.}^{\rm DBM}[\bm{x}, \tau | \bm{0}, 0] \;,
\end{equation}
{where ${\rm Prob.}^{\rm RDBM}[\bm{x} | r, \bm{x}_{\rm reset} = \bm{0}] $ denotes the probability for a RDBM to be at $\bm{x}$ in the NESS while resetting to $\bm{0}$ with rate $r$.}
From Eq. (\ref{eq:dbm-vicious-link}) we see that when $\bm{x_0} = \bm{0}$ the right hand side seems undefined. In reality, the divergence of the denominator will be regularized by the normalization constant in the VBM propagator. Hence, taking $\bm{x_0} = \epsilon \bm{a}$, where $a_{i} = -1 + 2 i /N$, we can re-write Eq. (\ref{eq:resetting-free-link}) as
\begin{equation} \label{eq:RDBM-VBM-renewal}
{\rm Prob.}^{\rm RDBM}_{\beta = 2, \mu=0}[\bm{x} | r, \bm{x}_{\rm reset} = \bm{0}] = \lim_{\epsilon \to 0^+} \frac{r}{\epsilon} \frac{\prod_{i < j} (x_i - x_j)}{ \prod_{i < j} (a_i - a_j) } \int_0^{+\infty} \dd \tau \; e^{-r \tau} {\rm Prob.}^{\rm VBM}[\bm{x}, \tau | \epsilon \bm{a}, 0] \;.
\end{equation}
From the same renewal arguments we know that the integral is the propagator of Resetting Vicious Brownian Motions (RVBM) which do not reset to the origin but to $\epsilon \bm{a}$, hence
\begin{equation} \label{eq:RDBM-RVBM-link}
{\rm Prob.}^{\rm RDBM}_{\beta = 2,\mu=0}[\bm{x} | r, \bm{x}_{\rm reset} = \bm{0}] = \lim_{\epsilon \to 0^+} \frac{1}{\epsilon} \frac{\prod_{i < j} (x_i - x_j)}{ \prod_{i < j} (a_i - a_j) } {\rm Prob.}^{\rm RVBM}[\bm{x} | r, \bm{x}_{\rm reset} = \epsilon \bm{a}] \;.
\end{equation}
Therefore, in the following sections, all results derived for the un-trapped RDBM, i.e. $\mu = 0$, can be directly used to obtain the behavior of a gas of $N$ RVBM by setting $\beta = 2$ and applying the appropriate weight defined in Eq. (\ref{eq:RDBM-RVBM-link}).

\subsection{Main results} \label{sec:intro-main-results}

Before getting into the details of the derivations, it is useful to summarise our principal results in the large $N$ limit, for the four main observables in the NESS: the average density, the extreme value statistics, the gap (i.e., the spacing between consecutive eigenvalues) statistics and the full counting statistics in the interval $[-L,L]$.

\vspace*{0.3cm}
\noindent {\bf Average density.}
In Section \ref{sec:model-density} we show that, in the limit of large $N$, the average density in the NESS of $\beta$-RDBM takes the scaling form
\begin{equation}\label{eq:main-density}
\rho_N(x | r) \simeq \sqrt{\frac{\mu}{N D}} \; f\left( x \sqrt{\frac{\mu}{N D}} , \, \frac{\mu}{r} \right) \;,
\end{equation}
where the scaling function $f(z, \gamma)$ is symmetric around $z=0$, has a finite support over $z \in [-\sqrt{2}, \sqrt{2}]$ and is normalized to unity, i.e., $\int_{-\sqrt{2}}^{\sqrt{2}}f(z,\gamma)\, dz = 1$.  This scaling function is independent of $\beta$ and is given by
\begin{equation} \label{eq:main-f}
f(z, \gamma) = \frac{\Gamma\left( 1 + \frac{1}{2 \gamma} \right)}{\sqrt{2 \pi} \Gamma\left( \frac{3}{2}+\frac{1}{2 \gamma}\right)} \left( 1 - \frac{z^2}{2} \right)^{\frac{1 + \gamma}{2 \gamma}}\, \,_2\,{F}_1 \left[1, \frac{1}{2\gamma}, \frac{3}{2} + \frac{1}{2\gamma}, 1 -\frac{z^2}{2} \right] \; \quad, \quad z \in [-\sqrt{2},+\sqrt{2}] \;,
\end{equation}
where $\,_2\,{F}_1[a, b, c, z]$ is the hypergeometric function. For some special values of $\gamma$ the hypergeometric function simplifies, for example for $\gamma = 1$ and $\gamma = 1/2$ we obtain
\begin{equation} \label{eq:main-density-special}
f(z, \gamma = 1) = \frac{1}{\sqrt{2}} - \frac{|z|}{2} \mbox{~~and~~} f(z, \gamma = 1/2) = \frac{2}{\pi} \left( \sqrt{2 - z^2} - |z| \, {\rm ArcCos} \left[ \frac{|z|}{\sqrt{2}} \right] \right) \;.
\end{equation}
The scaling function $f(z, \gamma)$ has the following asymptotic behaviors
\begin{equation} \label{eq:main-f-asymp}
f(z, \gamma) \approx \begin{dcases}
A_0 - \frac{|z|}{2 \gamma} &\mbox{~~when~~} |z| \to 0 \\
A_{+} (\sqrt{2} - |z|)^{(1 + 1/\gamma)/2} &\mbox{~~when~~} |z| \to \sqrt{2}
\end{dcases}\;,
\end{equation}
where the amplitudes $A_0$ and $A_+$ are given by
\bea \label{A0Ap}
A_0 = \sqrt{\frac{2}{\pi}} \frac{\Gamma\left(1+\frac{1}{2\gamma}\right)}{\Gamma \left(\frac{1}{2}+\frac{1}{2 \gamma}\right)}  \quad, \quad A_+ = \frac{2^{(1+\gamma)/(4 \gamma)}}{\sqrt{2\pi}} \frac{\Gamma\left( 1 + \frac{1}{2 \gamma}\right)}{\Gamma\left(\frac{3}{2} + \frac{1}{2 \gamma} \right)} \;.
\eea
In Fig. \ref{fig:density}, the analytical results are compared to numerical simulations, finding an excellent agreement. We also derive explicitly the average density for the un-trapped RDBM, i.e., in the limit $\mu \to 0$ -- see Eqs. \eqref{eq:density0} and \eqref{eq:def-f0}.

\vspace{0.3cm}
\noindent {\bf Extreme value statistics.} We show that, in the large $N$ limit, the distribution of the maximal eigenvalue, i.e. the position of the rightmost particle of the RDBM $x_{\rm max} = \max\{x_1, x_2, \cdots, x_N\}$, is given by
\begin{equation} \label{eq:main-max}
{\rm Prob.}[x_{\rm max} = x | r] \simeq \sqrt{\frac{\mu}{N D}} \; g\left( x \sqrt{\frac{\mu}{N D}}, \, \frac{\mu}{r} \right) \;,
\end{equation}
where the scaling function $g(z, \gamma)$ is independent of $\beta$. Remarkably, the function $g(z, \gamma)$, supported over $z \in [0, \sqrt{2}]$ and normalized to unity, has a very simple form (for $\gamma > 0$) given by
\begin{equation} \label{eq:main-g}
g(z, \gamma) = \frac{1}{2\gamma} z \left(1 - \frac{z^2}{2}\right)^{\frac{1}{2\gamma} - 1} \quad, \quad z \in [0,\sqrt{2}] \;.
\end{equation}
We recall that in the absence of resetting, i.e., for $r=0$, the corresponding distribution of $x_{\max}$ is highly concentrated around the upper edge $z = \sqrt{2}$ and its width is of order $O(N^{-2/3})$ (in units of $z$, see the left panel of Fig. \ref{fig:sketch}). The distribution of $x_{\max}$, centered around $\sqrt{2}$ and scaled by $N^{-2/3}$, is known as the Tracy-Widom distribution, which has an infinite support. In contrast, when a nonzero resetting rate $r>0$ is switched on, the distribution of EVS changes drastically. The distribution now spreads over $[0,\sqrt{2}]$ and is characterized by the exact scaling function in Eq. (\ref{eq:main-g}). Fig. \ref{fig:max} provides a comparison between analytical results and numerical simulations, showing excellent agreement.  The results in the limit $\mu \to 0$ are given in Eqs. 
\eqref{eq:evs-mu0} and \eqref{eq:def-g0}.

\vspace{0.3cm}
\noindent {\bf Gap statistics.} As in the $r=0$ case, computing the gap statistics in the large $N$ limit is complicated for any $r>0$. Therefore, following the Wigner surmise approach for $r=0$, we compute the gap statistics for a $2$-particle RDBM process with $r>0$. We find that the gap $\Delta = |x_1 - x_2|$ of a two-particle RDBM is given by
{
\begin{equation} \label{eq:main-gap}
{\rm Prob.}[\Delta = s] = \sqrt{\frac{\mu \beta}{D}} \; h_\beta\left( s \sqrt{\frac{\mu \beta}{D}}, \, \frac{\mu}{r} \right) \;,
\end{equation}}%
where the normalized scaling function $h_\beta(u, \gamma)$ defined for $u > 0$ and $\gamma > 0$ is given by
\begin{eqnarray} \label{eq:main-h}
h_\beta(u, \gamma) = \frac{\pi \sec(\frac{\pi \beta}{2})}{4 \gamma \Gamma\left( \frac{1 + \beta}{2} \right)} \Bigg( \frac{2^{1 - \beta} u^\beta \, \Gamma\left(\frac{1}{2\gamma}\right)}{\Gamma\left(\frac{1 - (\beta - 1)\gamma}{2 \gamma}\right) \Gamma\left( \frac{1+\beta}{2}\right) } \,_1 {F}_1\left[ \frac{(1 + \beta)\gamma - 1}{2 \gamma}, \frac{1+\beta}{2}, - \frac{u^2}{4} \right] \nonumber\\
- \frac{u}{\Gamma\left( \frac{3-\beta}{2}\right) }  \,_1 {F}_1\left[ 1 - \frac{1}{2\gamma}, \frac{3 - \beta}{2}, - \frac{u^2}{4} \right] \Bigg) \;,
\end{eqnarray}
where $_1F_1(a,b,z)$ is the Kummer's confluent hypergeometric function. Unlike the density, the scaling function itself depends on Dyson's index $\beta$. Since $\beta$ mainly influences short range behaviors of the gas it makes sense that it shows up only for the gaps. In fact, the small $u$ behavior depends crucially on $\beta$ and we find
\begin{equation} \label{eq:main-h-asymptotics-small-z}
h_\beta(u, \gamma) \underset{u \to 0}{\approx} \begin{dcases}
\left( \frac{\Gamma\left(\frac{1 - \beta}{2}\right) \Gamma\left(1 + \frac{1}{2\gamma}\right) }{2^\beta \Gamma\left(\frac{1 + \beta}{2}\right) \Gamma\left( \frac{1}{2\gamma} + \frac{1 - \beta}{2} \right)} \right)\,u^\beta & \mbox{~~for~~} \beta < 1 \\
 - \frac{u \log u}{2 \gamma} & \mbox{~~for~~} \beta = 1 \\
\frac{u}{2 \gamma (\beta - 1)} &\mbox{~~for~~} \beta > 1
\end{dcases} \;.
\end{equation}
On the other hand, for large $u$ and any $\beta>0$, the scaling function behaves as
\begin{equation} \label{eq:main-h-asymptotics-large-z}
h_\beta(u, \gamma) \underset{u \to \infty}\approx \frac{\Gamma\left(1 + \frac{1}{2 \gamma}\right)}{\Gamma\left(\frac{1 + \beta}{2}\right)} \left(\frac{u}{2}\right)^{\beta - 1/\gamma} e^{-u^2/4} \;.
\end{equation}
We recall that this is the exact result for $N=2$. In Fig. \ref{fig:gaps} we compare this result with the scaled distribution of the gap for $N=1000$ and show indeed that the Wigner surmise remains a very good approximation, even for $r>0$. For the un-trapped gas, i.e., for $\mu =0$, the explicit results for the gap statistics are given in Eqs. \eqref{eq:gap-mu0} and \eqref{eq:h0}.

\vspace{0.3cm}
\noindent{\bf Full Counting Statistics.} 
Finally, the distribution of the number $N_L$ of particles in a box $[-L, L]$ around the origin, in the limit of large $N$, satisfies a scaling form
\begin{equation} \label{main-fcs}
{\rm Prob.}[N_L=M | r] \simeq \frac{1}{N} \; q\left(\frac{M}{N}, \, L \sqrt{\frac{\mu}{N D}}, \, \frac{\mu}{r} \right) \;,
\end{equation}
where the scaling function $q(\kappa, \ell, \gamma)$ has a finite support over $\kappa \in \left[ \kappa_{\rm min} , 1\right]$ where
\begin{equation} \label{main-zmin}
\kappa_{\rm min} = \frac{1}{\pi} \ell \sqrt{2 - \ell^2} + \frac{2}{\pi} {\rm ArcTan}\left( \frac{\ell}{\sqrt{2 - \ell^2}} \right) \;.
\end{equation}
The scaling function, normalized to unity, i.e., $\int_{\kappa_{\min}}^1 q(\kappa,\ell,\gamma)\, \dd \kappa = 1$, and independent of $\beta$, is given by
\begin{align} \label{eq:main-d}
q(\kappa, \ell, \gamma) = \delta(\kappa - 1) \left[1 - \left(1 - \frac{\ell^2}{2}\right)^{\frac{1}{2 \gamma}}\right] + \frac{\pi \ell^2}{2 \sqrt{2} \, \gamma \, v_{\star}^3} \frac{1}{\sqrt{1-\frac{v_{\star}^2}{2}}} \left(1 - \frac{\ell^2}{v_{\star}^2} \right)^{\frac{1}{2 \gamma}-1} \;,
\end{align}
where $v_\star$ is the root of the following equation
\begin{equation} \label{eq:main-v-eq}
\frac{1}{\pi}\frac{\ell}{\sqrt{1 - v^{2 \gamma}}} \sqrt{2 - \frac{\ell^2}{1 - v^{2 \gamma}}} + \frac{2}{\pi} {\rm ArcTan}\left[ \frac{\frac{\ell}{\sqrt{1 - v^{2 \gamma}}}}{\sqrt{2 - \frac{\ell^2}{1 - v^{2 \gamma}}}} \right] = \kappa\;.
\end{equation}
The scaling function $q(\kappa, \ell, \gamma)$ has the asymptotic behaviors
\begin{eqnarray} \label{eq:main-left-asymptotic-d}
q(\kappa, \ell, \gamma) \approx
\begin{cases}
& \dfrac{1}{2\gamma} \left( \dfrac{\pi}{\ell\sqrt{2 - \ell^2}} \right)^{\frac{1}{2\gamma}} (\kappa-\kappa_{\min})^{\frac{1}{2\gamma} - 1} \quad, \quad\quad\quad  \hspace*{4.3cm}\kappa \to \kappa_{\rm min} \;,\\
& \\
& \delta(\kappa - 1) \left[1 - \left(1 - \frac{\ell^2}{2}\right)^{\frac{1}{2 \gamma}}\right] + \dfrac{\pi^{2/3}}{4 \gamma \, 6^{1/3}} \ell^2 \left(1 - \frac{\ell^2}{2}\right)^{\frac{1}{2 \gamma} - 1} \left({1 - \kappa}\right)^{-1/3} \quad, \quad \quad \kappa \to  1 \;. \\
\end{cases} 
\end{eqnarray}
%
In Fig. \ref{fig:fcs}, we provide a comparison between this analytical prediction and numerical simulations, showing a very good agreement. We also compute the $\mu \to 0$ limit of the FCS and the result is given in Eq.~\eqref{eq:free-q}.

\section{Derivation of the results} \label{sec:model}

In this section, we derive the analytical results in the large $N$ limit for the various observables described before, starting  
from Eqs. \eqref{eq:eigenvalues-jpdf-E} and \eqref{eq:def-Z-E}. We will then compare our analytical results with numerical
simulations performed by directly sampling the set of points $\{x_1, x_2, \cdots, x_N \}$ according to the joint distribution in 
the NESS in Eqs. \eqref{eq:eigenvalues-jpdf-E} and \eqref{eq:def-Z-E}. To carry out this procedure for general $\beta > 0$, 
it is convenient to use the so called the Dumitriu-Edelman procedure \cite{DE02}. Dumitriu and Edelman showed that
the eigenvalues of a tri-diagonal random matrix, with off-diagonal entries distributed as chi-square variables and 
parametrized by $\beta$ (with each entry scaled by a factor $\sigma(\tau)$), are distributed according to $\propto e^{-\beta E[x_1,\cdots,x_N;\tau]}$ where $E[x_1,\cdots,x_N;\tau]$
is given in Eq. (\ref{eq:def-Z-E}) with $\sigma^2(\tau)$ given in Eq. (\ref{eq:def-sigma}). For each fixed $\tau$, this generates $N$ eigenvalues
$\{x_1, \cdots, x_N\}$. Our simulation procedure consists of two steps: (i) we draw an exponentially distributed time $\tau$ from the distribution $p(\tau) = r e^{-r \tau}$ and for this chosen $\tau$ we get the $N$ eigenvalues of the tri-diagonal matrix and (ii) we compute any observable of interest for fixed $\tau$ and
finally average all possible $\tau$'s drawn from $p(\tau) = r e^{-r \tau}$.

\subsection{Density} \label{sec:model-density}

\begin{figure}[t]
\centering
\includegraphics[width=0.48\textwidth]{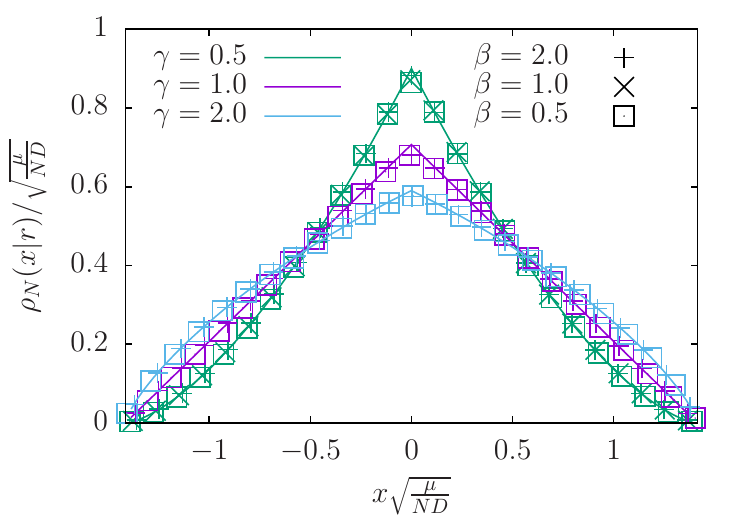} \hfill
\includegraphics[width=0.48\textwidth]{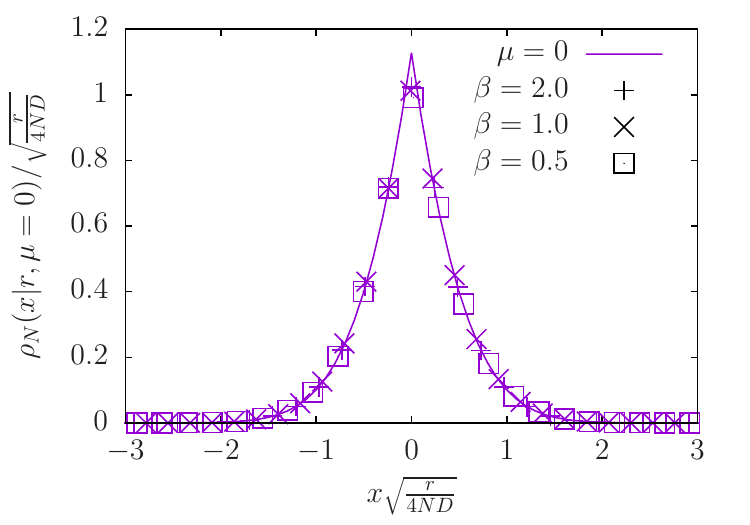}
\caption{{\bf Left.} Plot of the scaling function $f(z, \gamma)$ given in Eq. (\ref{eq:main-density}) vs the rescaled length $z=x \sqrt{\mu/(ND)}$, describing the average density profile of the RDBM in the NESS. The lines are the analytical prediction given by Eq. (\ref{eq:main-f}) and the points are the results of numerical direct sampling simulations as described in the introduction of Section \ref{sec:model}. Different colors correspond to different values of $\gamma = \mu/r$ (0.5, 1 and 2 for green, purple and blue respectively) and different symbols correspond to different values of $\beta =  0.5, 1, 2$. {\bf Right.} Plot of the scaling function $f_0(y)$ defined in Eq. (\ref{eq:def-f0}) vs the rescaled length $y = x \sqrt{r/(4ND)}$, describing the density of the untrapped ($\mu = 0$) RDBM in the NESS. The line corresponds to the theoretical prediction given in Eq. (\ref{eq:def-f0}) and the points are the results of numerical direct sampling simulations as described in the introduction of Section \ref{sec:model}. Different symbols correspond to different values of $\beta = 0.5, 1, 2$.}\label{fig:density}
\end{figure}

We start by studying the average spectral density in the NESS, which is defined as
\begin{equation} \label{eq:def-rho}
\rho_N(x | r) = \left\langle \frac{1}{N} \sum_{i = 1}^N \delta(x - x_i) \right\rangle_{\{x_i\} \sim {\rm Prob.}[x_1, \cdots, x_N | r]} \;,
\end{equation}
where the subscript outside the expectation denotes that the $x_i$'s are drawn from the joint distribution ${\rm Prob.}[x_1, \cdots, x_N | r]$ in 
Eqs. \eqref{eq:eigenvalues-jpdf-E} and \eqref{eq:def-Z-E}. We then get
\begin{equation} \label{eq:renewal-density}
\rho_N(x | r) = r \int_0^{+\infty} \dd \tau \; e^{-r \tau} \left\langle \frac{1}{N} \sum_{i = 1}^N \delta(x - x_i) \right\rangle_{\{x_i\} \sim {\rm Prob.}[x_1, \cdots, x_N | r = 0, t = \tau]} = r \int_0^{+\infty} \dd \tau \; e^{-r \tau} \rho_N(x | r = 0, \tau) \;,
\end{equation}
where $\rho_N(x | r = 0, \tau)$ is the average density of particles in the $\beta$-DBM at time $\tau$ without resetting. The latter is well known to be of the Wigner semi-circular form in the large $N$ limit and is independent of $\beta>0$~\cite{M91, PB20}. Using this result and the explicit expression $\sigma^2(\tau) = D(1-e^{-2\mu \tau})/\mu$, we get
{
\begin{equation} \label{eq:free-density}
\rho_N(x | r = 0, \tau) \simeq \frac{\sqrt{2 N \sigma^2(\tau) - x^2}}{\pi N \sigma^2(\tau)} = \sqrt{\frac{2 \mu}{N D (1 - e^{- 2 \mu \tau })\pi^2}\left(1 - \frac{\mu x^2}{ 2 N D ( 1 - e^{-2 \mu \tau} ) }\right)} \;.
\end{equation}
}
Substituting this result in Eq. (\ref{eq:renewal-density}) we obtain the large $N$ scaling form
\begin{equation}\label{eq:density-scaling}
\rho_N(x | r) \simeq \sqrt{\frac{\mu}{N D}} \; f\left(x \sqrt{\frac{\mu}{N D}}, \frac{\mu}{r}\right) \;,
\end{equation}
where the normalized scaling function $f(z, \gamma)$ supported over $z \in [-\sqrt{2}, \sqrt{2}]$ and for $\gamma > 0$ is given by
\begin{equation} \label{eq:def-f}
f(z, \gamma) = \frac{1}{2 \gamma} \int_{\log\frac{2}{2 - z^2}}^{+\infty} \dd u \; e^{- u/(2 \gamma)} \sqrt{\frac{2}{\pi^2 ( 1 - e^{-u} )}\left( 1 - \frac{z^2}{2 (1 - e^{-u})} \right)} \;.
\end{equation}
This integral can be performed explicitly leading to the result stated in Eq. (\ref{eq:main-f}). This scaling function is shown in Fig.~\ref{fig:density} for different values of $\gamma$ and is in perfect agreement with the numerical simulations. The asymptotic behaviors of this scaling function are given in Eqs. (\ref{eq:main-f-asymp}) and (\ref{A0Ap}). It is clear that the shape of this average density profile is drastically different from that of the Wigner semi-circular law. First it has a cusp as $z \to 0$, which results from this simultaneous resetting of the particles to $x=0$: this is quite different from the smooth quadratic behavior in the semi-circular form valid for $r=0$. Secondly, when $|z|$ approaches the edge of the support $z = \pm \sqrt{2}$, the spectral density vanishes with an exponent $(1+1/\gamma)/2$ larger than $1/2$ valid for $r=0$. In the limit when $r \to 0$, i.e. $\gamma = \mu/r \to \infty$, our spectral density reduces to the Wigner semi-circular law. To see this, we note that the integral in Eq. (\ref{eq:def-f}) is dominated by values of $u \gtrsim 2 \gamma$ for which $e^{-u} \ll 1$, recovering the usual RMT semi-circle law
\begin{equation} \label{eq:f-inf}
f(z, \gamma) \stackrel{\gamma \to +\infty}{\longrightarrow} \sqrt{\frac{2}{\pi^2}\left( 1 - \frac{z^2}{2}\right)}\; \frac{1}{2 \gamma} \int_{\log \frac{2}{2 - z^2}}^{+\infty} e^{-u/(2 \gamma)} \, \dd u = \sqrt{\frac{2}{\pi^2}\left( 1 - \frac{z^2}{2}\right)^{1 + 1/\gamma}} \stackrel{\gamma \to +\infty}{\longrightarrow} \sqrt{\frac{2}{\pi^2}\left( 1 - \frac{z^2}{2}\right)} \;.
\end{equation}

\vspace*{0.3cm}
\noindent{\bf The limit $\mu \to 0$.} The competition between the long-range attraction from resetting and the long-range repulsion from the logarithmic interaction balances out even in the absence of a confining potential. Hence, it is possible to reach a new NESS with Dyson's logarithmic repulsion combined with resetting. To obtain this limit, we consider the scaling form of the spectral density in Eq. (\ref{eq:density-scaling}). When $\mu \to 0$, the ratio $\gamma = \mu/r \to 0$ and also the scaled distance $z = x \sqrt{\mu/(N\,D)} \to 0$. Since we want to find the density at a fixed position $x$, we need to take the simultaneous limits $\gamma \to 0$, $z \to 0$, but with the ratio $z^2 / \gamma = x^2 r / (N D)$ fixed. Taking these limits in Eq. (\ref{eq:density-scaling}) and Eq. (\ref{eq:def-f}) we get
\begin{equation} \label{eq:density0}
\rho_N(x | r) \stackrel{\mu \to 0}{\longrightarrow} \sqrt{\frac{r}{4 N D}} f_0 \left( x \sqrt{\frac{r}{4 N D}} \right) \;,
\end{equation}
where 
\begin{equation} \label{eq:def-f0}
f_0(y) = \frac{2}{\pi} \int_{y^2}^{+\infty} \dd v \; e^{-v} \sqrt{\frac{v - y^2}{v^2}} = \frac{2 e^{-y^2}}{\sqrt{\pi}} - 2 |y| \, {\rm erfc} |y| \;,
\end{equation}
is the normalized scaling function which is now defined on the unbounded support $y \in \mathbb{R}$. It is plotted in the right panel of Fig. \ref{fig:density} and is in perfect agreement with the numerical simulations. The asymptotics of this scaling function $f_0(y)$ are given by
\begin{equation} \label{eq:f0-asymptotics}
f_0(y) \approx \begin{dcases}
\frac{2}{\sqrt{\pi}} - 2 |y| &\mbox{~~as~~} |y| \to 0\\
\frac{e^{-y^2}}{y^2 \sqrt{\pi}} &\mbox{~~as~~} |y| \to \infty
\end{dcases} \;.
\end{equation}
In the absence of a confining trap, the density is now defined over the full real line instead of a bounded region around the origin, but the resetting forms a cusp at the origin, concentrating the measure there, which allows the gas to reach a NESS with an unbounded support.

\subsection{EVS} \label{sec:model-evs}

We now turn our attention to the EVS, i.e., to the study of the distribution of the position $x_{\rm max} = \max \{x_1, \cdots, x_N\}$ of the rightmost particle in the RDBM. In the absence of resetting $r=0$, it is well known that $x_{\max}$, appropriately centered and scaled in the limit of large $N$, is distributed via the 
Tracy-Widom law \cite{F10,TW94,TW96}. More precisely, $x_{\max}$, as a random variable, can be expressed, for large $N$, as  
\begin{equation}\label{eq:free-max}
x_{\rm max}(\tau, r = 0) \approx \sqrt{2 N \sigma(\tau)^2} + \frac{\sigma(\tau)}{\sqrt{2}} N^{-1/6} \chi_\beta = \sqrt{\frac{D (1 - e^{-2 \mu \tau})}{\mu}} \left[ \sqrt{2 N} + \frac{1}{\sqrt{2}} N^{-1/6} \chi_\beta\right] \;,
\end{equation}
where $\chi_\beta$ is an $\mathcal{O}(1)$ random variable whose law is given by the Tracy-Widom distribution of index $\beta$. Consequently the scaled maximum can be expressed as 
\bea \label{scaled_xmax}
\sqrt{\frac{\mu}{N\, D}} x_{\max}(\tau, r= 0)\approx \sqrt{(1 - e^{-2 \mu \tau})} \left[ \sqrt{2} + \frac{1}{\sqrt{2}} N^{-2/3}\, \chi_\beta\right] \;.
\eea
Since the width of the distribution of $x_{\max}$ scales as $N^{-1/6}$ in Eq. (\ref{eq:free-max}) (or equivalently as $O(N^{-2/3})$ for the scaled maximum in Eq. (\ref{scaled_xmax})), it follows that, for large $N$, the distribution gets more and more concentrated around the mean $\sqrt{2 N \sigma^2(\tau)}$. When the simultaneous resetting $r$ is switched on, using Eqs. \eqref{eq:eigenvalues-jpdf-E} and \eqref{eq:def-Z-E}, 
the probability distribution can be written as  
\begin{equation} \label{eq:full-renewal-evs}
{\rm Prob.}[x_{\rm max} = x | r] = r \int_0^{+\infty} \dd \tau \; e^{-r \tau}\; {\rm Prob.}[x_{\rm max}(\tau,r=0) = x] \;,
\end{equation}
where the random variable $x_{\rm max}(\tau, r = 0)$ is distributed via Eq. (\ref{eq:free-max}). In the large $N$ limit, it turns out that the dominant contribution to this integral comes from the deterministic part in Eq. (\ref{eq:free-max}). The stochastic part containing the Tracy-Widom variable gives only subleading contributions for all $r>0$. Hence one can approximate Eq. (\ref{eq:full-renewal-evs}) by
\begin{equation} \label{eq:full-renewal-evs-delta}
{\rm Prob.}[x_{\rm max} | r] \approx r \int_0^{+\infty} \dd \tau \; e^{-r \tau}\; \delta\left[x_{\rm max} - \sqrt{\frac{2 N D(1 - e^{-2\mu \tau})}{\mu}}\right] \;.
\end{equation} 
}
This integral can be done explicitly and has a scaling form
\begin{equation}\label{eq:max-scaling}
{\rm Prob.}[x_{\rm max} = x | r] \simeq \sqrt{\frac{\mu}{N D}} \; g\left( x \sqrt{\frac{\mu}{N D}}  , \frac{\mu}{r} \right) \;,
\end{equation}
where the normalized scaling function $g(z, \gamma)$ is supported over $z \in [0, \sqrt{2}]$ (strictly for $\gamma > 0$) by
\begin{equation} \label{eq:g-def}
g(z, \gamma) = \frac{1}{2\gamma} z \left(1 - \frac{z^2}{2}\right)^{\frac{1}{2\gamma} - 1} \quad, \quad {\rm with} \quad \gamma = \frac{\mu}{r} \;. 
\end{equation}
This scaling function is shown in Fig. \ref{fig:max} and is in perfect agreement with numerical simulations. Notice that when $\gamma \to +\infty$ the distribution concentrates around the upper limit $z=\sqrt{2}$ of the support recovering the deterministic part of Eq. (\ref{eq:free-max}). However, for any $\gamma >0$, the maximal eigenvalue has a nontrivial distribution parametrized by $\gamma$. This distribution diverges as $z \to \sqrt{2}$ for $\gamma>1/2$, while it vanishes  as $z \to \sqrt{2}$ for $\gamma <1/2$, thus displaying a ``shape transition'' at $\gamma=1/2$. Once again, this is extremely different from the behavior of the EVS for the Dyson log gas. In Dyson log gas, the rightmost particle is always at the right edge of the support of the density, i.e., around the scaled distance $z = \sqrt{2}$, with small fluctuations of order $O(N^{-2/3})$. The centered and the scaled distribution is described by the Tracy-Widom law. On the other hand, in our gas, the 
rightmost particle can be anywhere between $z=0$ and $z= \sqrt{2}$ and its distribution has a width of order $O(1)$ in the large $N$ limit. This distribution may either diverge or vanish as $z \to \sqrt{2}$, depending on whether $\gamma > 1/2$ or $\gamma<1/2$. Thus, as one switches on resetting, the whole distribution of $x_{\max}$ is shifted from the upper edge $\sqrt{2}$ towards the trap center $z=0$, in stark contrast with the $r=0$ case. 
This is a clear indication of the strong attractive correlations generated by resetting. Indeed, in Dyson log gas the strong repulsive correlations ensure that the particles will always try to spread as much as possible within the trap, therefore pushing the rightmost particle to the right edge of the support. In our gas, the strong attractive correlations balance the repulsive ones and hence allow for the rightmost particle to not necessarily concentrate near the upper edge $z=\sqrt{2}$ of the support.

\vspace*{0.3cm}
\noindent{\bf The limit $\mu \to 0$.} As was done previously for the density, to obtain the result for the untrapped gas, i.e. $\mu \to 0$, we have to take $\gamma \to 0$ and $z \to 0$ keeping $z^2 / \gamma = x^2 r  / (N D)$ fixed. Taking these limits in Eq. (\ref{eq:max-scaling}) and Eq. (\ref{eq:g-def}) we obtain in the large $N$ limit
\begin{equation} \label{eq:evs-mu0}
{\rm Prob.}[x_{\rm max} = x | r] \stackrel{\mu \to 0}{\longrightarrow} \sqrt{\frac{r}{4 N D}} \; g_0\left(x \sqrt{\frac{r}{4 N D}} \right) 
\end{equation}
where
\begin{equation} \label{eq:def-g0}
g_0\left(y \right) = 2 \, y \, e^{-y^2} \;,
\end{equation}
is the normalized scaling function defined on $y > 0$. Interestingly, the same scaling function $g_0(y)$ appears in the problem of $N$ simultaneously resetting noninteracting Brownian motions~\cite{BLMS23}. In that problem it was shown that the distribution of $x_{\max}$ exhibits a scaling form 
 \begin{equation} \label{eq:evs_inde}
{\rm Prob.}[x_{\rm max} = x | r] \approx \sqrt{\frac{r}{4  D \ln N}} \; g_0\left(x \sqrt{\frac{r}{4 D \ln N}} \right) \;,
\end{equation}
where $g_0(y)$ is given in (\ref{eq:def-g0}). Hence both problems exhibit exactly the same scaling function but the $N$-dependence of the 
scale of $x_{\max}$ is different. In the former case it scales as $\sqrt{N}$, while in the latter case it scales as $\sqrt{\ln N}$. The reason behind the appearance
of the same scaling function $g_0(y)$ can be understood qualitatively as follows. In the RDBM with $\mu = 0$, since the 
particles are now untrapped, diffusion will make the semi-circle density spread like $\sim \sqrt{t}$. The logarithmic repulsion does not change the diffusive spread but just renormalises $\sqrt{t}$ to $\sqrt{N \, t}$. Thus the deterministic part of the maximum behaves as $\sqrt{N\,t}$, which gives rises to the scaling function $g_0(y)$ as shown above. In the case of noninteracting Brownian particles with simultaneous resetting, the deterministic part of $x_{\max}$ behaves as $\sqrt{(\ln N)\,t}$. While the fluctuating parts are very different in the two problems, the deterministic part scales as $\sqrt{t}$ in both problems, albeit with different $N$ dependent prefactors. When the resetting is switched on, the dominant contribution to the distribution of $x_{\max}$ comes only from the deterministic part,   
while the stochastic part contributes to subleading behaviors. Since both problems share the same deterministic time dependent evolution, the leading large $N$ behavior in the presence of resetting is exactly the same in both problems, up to the $N$ dependent rescaling factor.

\begin{figure}[t]
\centering
\includegraphics[width=0.48\textwidth]{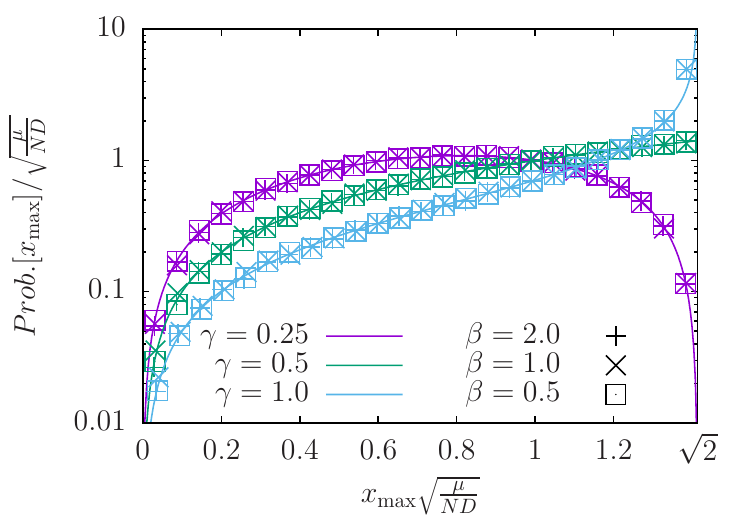}\hfill
\includegraphics[width=0.48\textwidth]{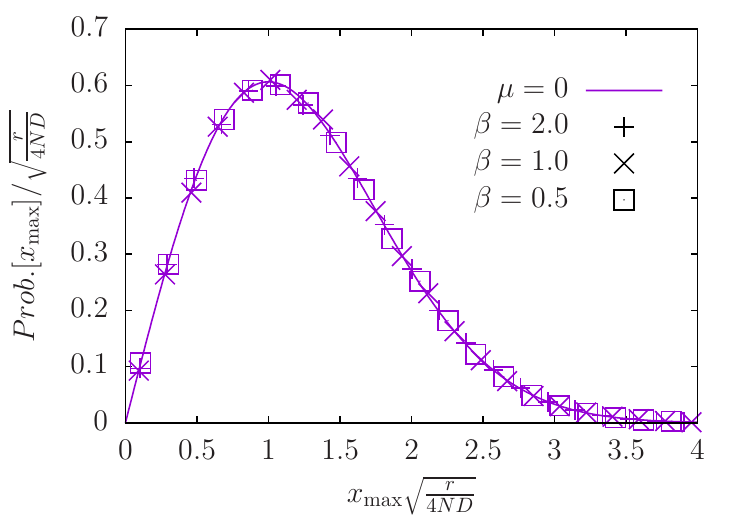}
\caption{{\bf Left.} Plot of the scaling function $g(z, \gamma)$ defined in Eq. (\ref{eq:g-def}) vs the rescaled maximum $z =x_{\max} \sqrt{\mu/(ND)}$, describing the probability distribution function (PDF) of $x_{\max}$ representing the position of the rightmost particle of the RDBM in its NESS. The lines are the analytical prediction given in Eq. (\ref{eq:g-def}) and the points are the results of numerical direct sampling simulations as described in the introduction of Section \ref{sec:model}. Different colors correspond to different values of $\gamma$ (0.25, 0.5 and 1 for purple, green and blue) and different symbols corresponds to different values of $\beta = 0.5,1,2$. {\bf Right.} Plot of the scaling function $g_0(y)$ defined in Eq. (\ref{eq:def-g0}) vs the rescaled maximum $y=x_{\max}\sqrt{r/(4 ND)}$ describing the PDF of the rightmost particle's position of the untrapped ($\mu = 0$) RDBM in its NESS. The line is the analytical prediction given by Eq. (\ref{eq:def-g0}) and the crosses are the results of numerical direct sampling simulations as described in the introduction of Section \ref{sec:model}. Different symbols correspond to different values of $\beta=0.5,1,2$.}\label{fig:max}
\end{figure}

\subsection{Gap statistics} \label{sec:model-gap}

We now turn our attention to the spacings in the gas. The simplest statistical description of the spacing in RMT is given by Wigner's surmise. Wigner computed exactly the statistics of the gap $s = |x_1 - x_2|$ for a 2 by 2 matrix and conjectured that it would be a good approximation for the mean level spacing for matrices of any size. Numerical results have shown that it is indeed the case~\cite{M91}. We will follow a similar procedure here in the presence of resetting $r>0$. Our starting point then is the joint distribution of eigenvalues in Eq. (\ref{eq:eigenvalues-jpdf}) which reads, for $N=2$, 
\begin{equation} \label{eq:eigenvalues-jpdf_N2}
{\rm Prob.}[x_1, x_2 | r ] = A_\beta \int_0^{+\infty} r \dd\tau\; e^{-r\tau} \frac{1}{\sigma(\tau)^{\beta + 2}} \exp[ -\frac{\beta}{2\sigma^2(\tau)} (x_1^2 + x_2^2) ] \, |x_1-x_2|^\beta\;,
\end{equation}
where $\sigma^2(\tau) = D(1-e^{-2\mu \tau})/\mu$ is given in Eq. (\ref{eq:def-sigma}) and the normalisation constant $A_\beta$ is given by
\bea \label{Ab}
A_\beta = \frac{2}{\sqrt{\pi}}\left(\frac{\sqrt{\beta}}{2} \right)^{\beta + 2} \frac{1}{\Gamma\left( \frac{1+\beta}{2}\right)} \;.
\eea 
The distribution of the spacing $\Delta = |x_1 - x_2|$ is then given by
\bea \label{p_gap1}
{\rm Prob.}[\Delta = s \vert r] = \int {\rm Prob.}[x_1, x_2 | r ] \; \delta \left[ |x_1-x_2| = s\right] \; \dd x_1 \dd x_2 \;,
\eea
where ${\rm Prob.}[x_1, x_2 | r ]$ is given in Eq. (\ref{eq:eigenvalues-jpdf_N2}). This double integral can be easily done by making the change of variables $x_1+x_2 = y_1$ and $x_1 - x_2 = y_2$. Performing these integrals we get  
{
\begin{equation} \label{eq:renewal-s}
{\rm Prob.}[\Delta = s | r] = \frac{r \sqrt{\beta} }{2^\beta \Gamma\left(\frac{1 + \beta}{2} \right)} \left(s \sqrt{\beta} \right)^\beta  \int_0^{+\infty} \dd \tau \; \frac{e^{-r \tau}}{\sigma(\tau)^{1 + \beta}}  \exp[ - \frac{s^2}{4} \frac{\beta}{\sigma^2(\tau)} ]  \;.
\end{equation}
}
This can be written in a scaling form as
{
\begin{equation} \label{eq:s-scaling}
{\rm Prob.}[\Delta =s \vert r] = \sqrt{\frac{\mu \beta}{D}} h_\beta\left( s \sqrt{\frac{\mu \beta}{D}}, \frac{\mu}{r}  \right) \;,
\end{equation}
}
where 
\begin{equation} \label{eq:h-def}
h_\beta(u, \gamma) = \frac{u^\beta}{2^\beta \Gamma\left(\frac{1 + \beta}{2} \right)} \int_0^{+\infty} \dd T \; e^{-T} \left(\frac{1}{1 - e^{-2\gamma T}}\right)^{\frac{1+\beta}{2}} \exp[ - \frac{u^2}{4(1 - e^{-2\gamma T})}] \;.
\end{equation}
Making the change of variable $v = 1 - e^{-2 \gamma T}$ we can compute this integral explicitly which then gives the scaling function $h_\beta(u,\gamma)$ explicitly as given in Eq. (\ref{eq:main-h}), together with its asymptotic behaviors as $u \to 0$ and $u \to \infty$ given in Eqs. (\ref{eq:main-h-asymptotics-small-z}) and (\ref{eq:main-h-asymptotics-large-z}).

\begin{figure}[t]
\centering
\includegraphics[width=0.32\textwidth]{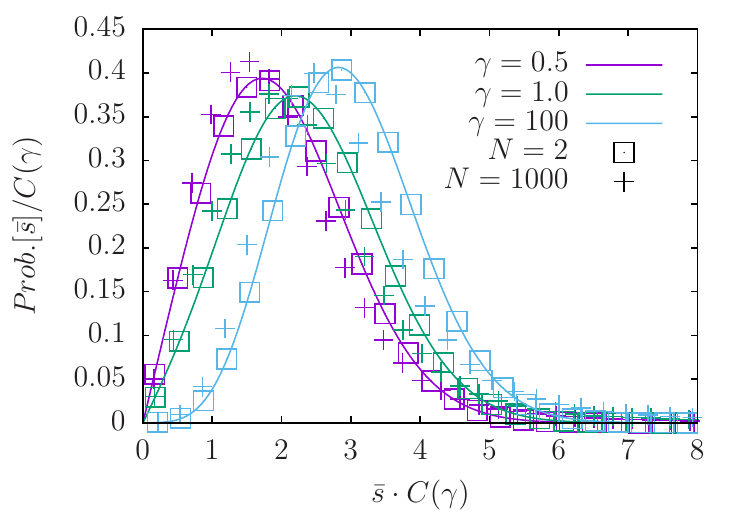}
\hfill
\includegraphics[width=0.32\textwidth]{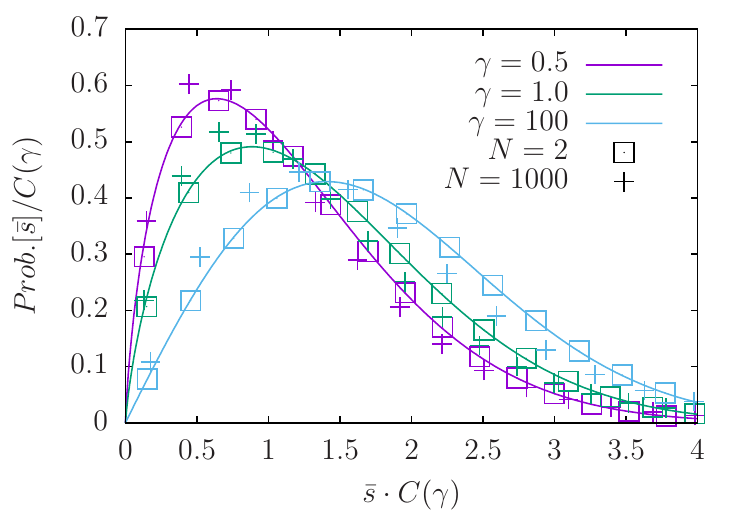}
\hfill
\includegraphics[width=0.32\textwidth]{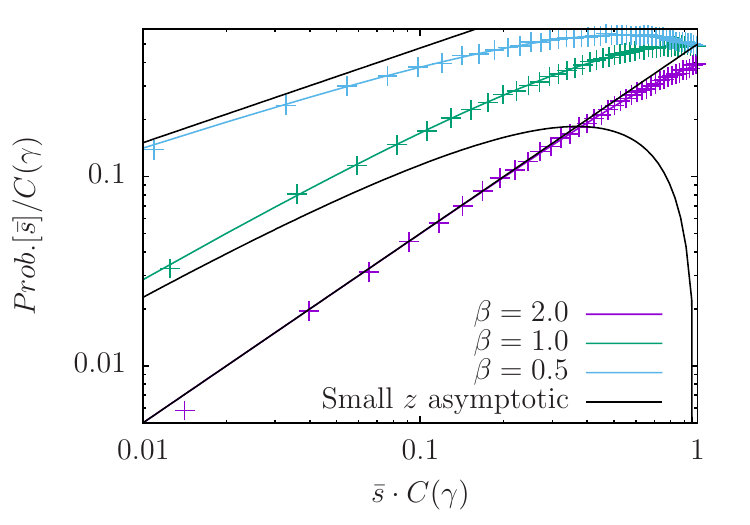}
\caption{{\bf Left panel:} Plot of ${\rm Prob.}[\bar{s}\vert r]$ in Eq. \eqref{eq:s-mean-C} as a function of the scaled distance $\bar{s}\,C(\gamma)$ with $C(\gamma)$ given in Eq. (\ref{eq:def-Cgamma}) and the scaling function $h_{\beta}(u, \gamma)$ given in Eq. (\ref{eq:main-h}) describing the PDF of the gap of the $\beta$-RDBM for $\beta=4$ in its NESS with $N=2$. The lines correspond to the analytical prediction given by Eq. (\ref{eq:main-h}) and the points are the results from numerical direct sampling simulations as described in the introduction of Section \ref{sec:model} for $N=2$ and for $N=1000$. {\bf Middle panel:} the same as the left panel but with $\beta = 1$. The curves in
the left and middle panels confirm that the Wigner surmise is a good approximation even in the presence of resetting. Different colors correspond to different values of $\gamma$ (0.5, 1 and 100 for purple, green and blue) and different symbols correspond to $N=2$ and $N=1000$. {\bf Right panel:} 
it represents  a zoom of the small-$z$ asymptotics given in Eq. (\ref{eq:main-h-asymptotics-small-z}), where we fixed $\gamma = 1$ and the different colors correspond to $\beta = 0.5,1,2$.}\label{fig:gaps}
\end{figure}

\vspace*{0.3cm}
From Eq. (\ref{eq:s-scaling}) and Eq. (\ref{eq:h-def}) we can obtain the average gap explicitly, yielding
{
\begin{equation} \label{eq:s-mean}
\langle s \rangle = \sqrt{\frac{\pi D}{\mu \beta}} \frac{\Gamma\left(1 + \frac{\beta}{2}\right)\Gamma\left( 1 + \frac{1}{2\gamma} \right)}{\Gamma\left(\frac{1 + \beta}{2}\right) \Gamma\left( \frac{3}{2} + \frac{1}{2 \gamma} \right)}
\end{equation}
}
Hence the distribution of the scaled spacing $\bar{s} = s / \langle s \rangle$ is given by
\begin{equation} \label{eq:s-mean-C}
{\rm Prob.}[\bar{s}\vert r] = C(\gamma) \, h_\beta\left( \bar{s} \, C(\gamma), \, \frac{\mu}{r} \right) \;,
\end{equation}
where 
\begin{equation}\label{eq:def-Cgamma}
C(\gamma) = \sqrt{\pi} \frac{\Gamma\left(1 + \frac{\beta}{2}\right)\Gamma\left( 1 + \frac{1}{2\gamma} \right)}{\Gamma\left(\frac{1 + \beta}{2}\right) \Gamma\left( \frac{3}{2} + \frac{1}{2 \gamma} \right)} \;,
\end{equation}
and the scaling function $h_\beta(u,\gamma)$ is given in Eq. (\ref{eq:main-h}). Once again, for some special values of $\gamma$ the scaling function simplifies. Indeed for $\gamma = 1/2$ we have
\begin{equation} \label{eq:h-gamma-half}
h_\beta(u, 1/2) = \frac{u}{2} \frac{\Gamma\left(\frac{\beta - 1}{2}, \frac{u^2}{4}\right)}{\Gamma\left( \frac{1 + \beta}{2} \right)} \;,
\end{equation}
where $\Gamma(a, z)$ is the incomplete $\Gamma$-function. For $\gamma = 1$ and $\beta = 1$ we get
\begin{equation} \label{eq:h-gamma-1}
h_{\beta = 1}(u, 1) = \frac{u}{4} e^{-u^2/8} \, K_0(u^2 / 8) \;,
\end{equation}
where $K_0(z)$ is the modified Bessel function of index $0$. Furthermore, for $r \to 0$, i.e., for $\gamma \to +\infty$, we recover the original form of Wigner's surmise
\begin{equation} \label{eq:h-gamma-inf}
h_\beta(u, +\infty) = \frac{1}{2^\beta \Gamma\left( \frac{1 + \beta}{2} \right)} u^\beta e^{- u^2/4} \; .
\end{equation}
The scaling function in Eq. (\ref{eq:s-mean-C}) is plotted in Fig. \ref{fig:gaps} and is in perfect agreement with numerical simulations when $N = 2$. As is the case in RMT, i.e., for $r=0$, the $N = 2$ result remains a good approximation of the scaled spacing distribution for $N>2$. As can be seen in Fig. \ref{fig:gaps}, as one increases $r$, i.e., decreases $\gamma$, the peak of the distributions move towards smaller value of $\bar{s}$, indicating that the typical gap decreases with increasing $r$. This is consistent with the physical expectation that resetting generates effective attractive correlations between particles in the NESS and hence decrease the typical spacing between particles. This can also be seen from the small $u$ asymptotics in Eq. (\ref{eq:main-h-asymptotics-small-z}). As $u \to 0$, the gap distribution vanishes as $\sim u^\beta$ for $\beta < 1$, as  $-u\log u$ for $\beta = 1$ and linearly for $\beta > 1$. This is to be compared with the $r=0$ case \cite{M91,F10}, where the distribution vanishes as $u^\beta$ as $u \to 0$ for all $\beta>0$. Hence, for $r>0$, the small gaps have larger probability to occur, reflecting the attractive correlations generated by resetting.

\vspace*{0.3cm}
\noindent{\bf The limit $\mu \to 0$.} As in the previous cases, when we take the $\mu \to 0$ limit to study the gap distribution for the 
un-trapped gas, we see from Eq. (\ref{eq:s-scaling}) that we have to simultaneously take the limits $\gamma \to 0$, $u \to 0$ while keeping the ratio 
$u^2 / \gamma =  r s^2 \beta / D$ fixed in Eqs. (\ref{eq:s-scaling}) and (\ref{eq:h-def}). This yields
{
\begin{equation} \label{eq:gap-mu0}
{\rm Prob.}[\Delta = s | r] = \sqrt{\frac{r \beta}{2 D}} \; h_0\left( s \sqrt{\frac{r \beta}{2 D}} \right) \;,
\end{equation}
}
where
\begin{equation} \label{eq:h0}
h_0(y) =  \frac{y^{\frac{1 + \beta}{2}}}{2^{\frac{\beta - 1}{2}} \Gamma\left( \frac{1+\beta}{2} \right)} K_{\frac{\beta - 1}{2}}\left( y \right) \;,
\end{equation}
where $K_\nu(z)$ is the modified Bessel function of index $\nu$.  

\vspace*{0.3cm}
Similarly, in the untrapped limit $\mu \to 0$, the scaled PDF of the level-spacing, in units of mean level-spacing $\langle s \rangle$, is given by
\begin{equation} \label{eq:mean-gap-mu0}
{\rm Prob.}[\bar{s} \vert r] = \sqrt{\pi} \frac{\Gamma\left(1 + \frac{\beta}{2}\right)}{\Gamma\left( \frac{1 + \beta}{2} \right)} \, h_0\left( \sqrt{\pi} \frac{\Gamma\left(1 + \frac{\beta}{2}\right)}{\Gamma\left( \frac{1 + \beta}{2} \right)} y \right) \;.
\end{equation}
The asymptotic behavior of the scaling function $h_0(y)$ as $y \to 0$ is different for $\beta = 1$ and $\beta = 2,4$. One finds 
\begin{equation} \label{eq:h0-asymptotics-small-z}
h_0(y)  \underset{y \to 0}{\approx} \begin{dcases}
\left( \frac{\Gamma\left(\frac{1 - \beta}{2}\right)}{2^\beta \Gamma\left( \frac{1 + \beta}{2} \right)} \right)\,y^\beta \quad &{\rm for} \quad \beta < 1\\
- y \log y & {\rm for}\quad \beta = 1 \\
\frac{y}{\beta - 1} &{\rm for} \quad \beta >1 
\end{dcases} \;.
\end{equation}
The large-$y$ asymptotic behavior is given by
\begin{equation} \label{eq:h0-asymptotics-large-z}
h_0(y) \underset{y \to \infty}{\approx}  \sqrt{\frac{\pi}{2^\beta}} \frac{1}{\Gamma\left( \frac{1 + \beta}{2} \right)} y^{\beta/2} e^{-y} \;.
\end{equation}
Comparing Eq. (\ref{eq:h0-asymptotics-small-z}) and Eq. (\ref{eq:h0-asymptotics-large-z}) with Eq. (\ref{eq:main-h-asymptotics-small-z}) and Eq. (\ref{eq:main-h-asymptotics-large-z}) we see that untrapping the gas has not changed the small $y$ asymptotic behavior (apart from a prefactor) but the large $y$ asymptotics are significantly different: the Gaussian tail for $\mu > 0$ changes to a much slower exponential tail when $\mu \to 0$. Thus the right tail is significantly stretched in the absence of a confining potential. Physically this is what we would expect, since the gas is now un-trapped it can reach configurations with possibly much larger gaps.

\subsection{Full counting statistics} \label{sec:model-fcs}

In this section, we compute the distribution of $N_L$ that denotes the number of particles in the  interval $[-L, L]$ in the NESS induced by a nonzero resetting rate $r>0$. Clearly $N_L$ is a random variable that fluctuates from one sample to another sample of the gas. This random variable can be expressed as 
\bea \label{indic}
N_L = \sum_{i=1}^N {\mathbb I}_{x_i \in [-L,L]} \;,
\eea
where ${\mathbb I}_{x_i \in [-L,L]}$ is an indicator function, which takes the value ${\mathbb I}_{x_i \in [-L,L]} = 1$ if the $i$-th particle is inside the interval $[-L,L]$ and ${\mathbb I}_{x_i \in [-L,L]} = 0$ otherwise. Therefore the probability distribution of $N_L$ is given by
\bea \label{FCS_1}
{\rm Prob.}[N_L = M \vert r] = \int \dd x_1 \cdots \dd x_N \delta \left[ M - \sum_{i=1}^N {\mathbb I}_{x_i \in [-L,L]} \right] {\rm Prob.}[x_1, \cdots, x_N \vert r]
\eea
where $ {\rm Prob.}[x_1, \cdots, x_N \vert r]$ is given in Eq. (\ref{eq:eigenvalues-jpdf}). Interchanging the integrals over $\tau$ and the $x_i$'s one can re-write Eq.~(\ref{FCS_1}) as
\bea \label{FCS_2}
{\rm Prob.}[N_L = M \vert r] &=& r\int_0^\infty \dd \tau  e^{-r \tau} \int \dd x_1 \cdots \int \dd x_N \delta\left[M -  \sum_{i=1}^N {\mathbb I}_{x_i \in [-L,L]}\right]\nonumber \\
&\times& \frac{1}{Z_N(\beta)} \frac{1}{\sigma(\tau)^{N + \beta N(N-1)/2}} \exp[ -\frac{\beta}{2}\left(\frac{1}{\sigma^2(\tau)} \sum_{i = 1}^N x_i^2 - \sum_{i \neq j} \ln |x_i - x_j| \right)] \nonumber \\
&=&   r\int_0^\infty \dd \tau  \,e^{-r \tau} \, {\rm Prob.}[N_L(\tau) = M ; \tau] \;,
\eea  
where we have identified the integrals over the $x_i$'s as the probability distribution ${\rm Prob.}[N_L(\tau) = M ; \tau]$ of the number of particles $N_L(\tau)$ in the interval $[-L,L]$ in the Coulomb gas with energy in Eq. (\ref{eq:def-Z-E}) parametrized by $\tau$ that appears inside the variance $\sigma^2(\tau) = D(1-e^{-2 \mu \tau})/\mu$. This FCS ${\rm Prob.}[N_L(\tau) = M ; \tau]$ of the Coulomb gas has been studied in the RMT literature~\cite{DM63,CL95,FS95,MMSV14,CLM15,MMSV16}. In particular, the distribution has a peak around its mean $\langle N_L(\tau)\rangle = O(N)$ and it fluctuates around this mean over a scale $O(\sqrt{\ln N})$. In other words, the random variable $N_L(\tau)$ in the Coulomb gas with parameter $\tau$ can be expressed, for large $N$, as 
\bea \label{FCS_3}
N_L(\tau) \approx \langle N_L(\tau)\rangle + \sqrt{\ln N} \, W \;,
\eea
where $W$ is an $N$-independent, but $\ell$-dependent, random variable with magnitude of order $O(1)$. This is because the log-gas is very rigid: in fact, this is an example of hyperuniformity~\cite{T18} where the fluctuations scale less than linearly as $N \to \infty$. The mean value $ \langle N_L(\tau)\rangle$ can be computed explicitly limit as
\begin{equation} \label{eq:def-FCS}
\langle N_L(\tau) \rangle = N \int_{-L}^{L} \dd x \; \rho_N(x | r = 0, \tau) \;,
\end{equation}
where the average density $\rho_N(x | r = 0, \tau)$ of the Coulomb gas with parameter $\tau$ is given explicitly (for large $N$) by the Wigner semi-circular form 
in Eq. (\ref{eq:free-density}). It is convenient to first introduce the dimensionless length $\ell$ as
\begin{equation}\label{eq:def-ell}
\ell = L \sqrt{\frac{\mu}{N D}} \;.
\end{equation}
Since, for large $N$, the gas is confined on the interval $z \in [-\sqrt{2},+\sqrt{2}]$, we will henceforth consider only $\ell \in [0, \sqrt{2}]$. For $\ell > \sqrt{2}$, $N_L(\tau)$ saturates to $N$ with probability one and there are no fluctuations. Therefore FCS have a nontrivial behavior only for $\ell < \sqrt{2}$.

\begin{figure}[t]
\includegraphics[width = 0.55 \linewidth]{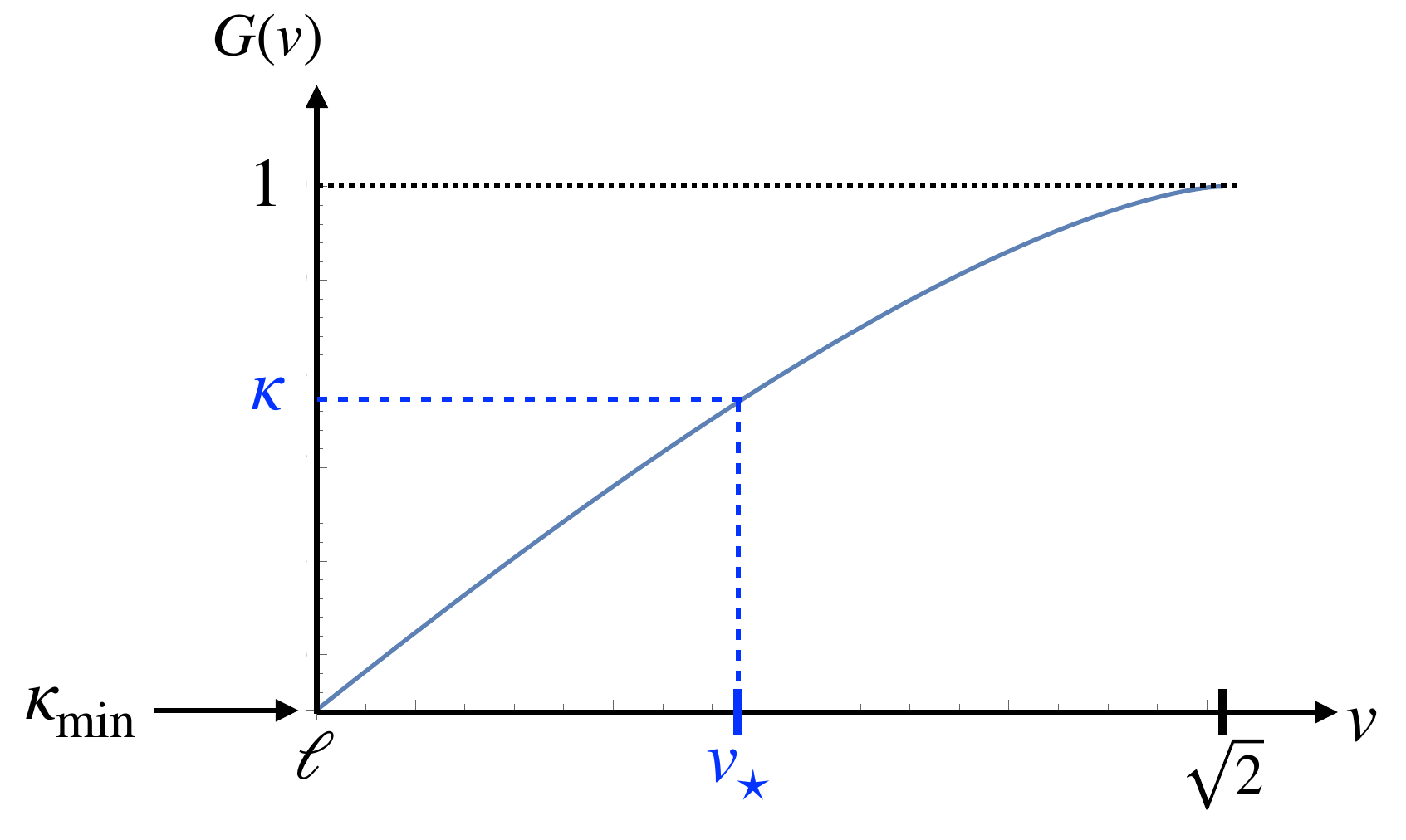}
\caption{Plot of the function $G(v)$ given in Eq. (\ref{def_G}). For a given $\kappa$, one can then read off $v_\star$ such that $G(v_\star) = \kappa$ as stated in Eq. (\ref{eq:v-eq}). The lowest allowed value $\kappa_{\min}$ is given in Eq. (\ref{eq:def-zmin}).}\label{Fig_G}
\end{figure}

\vspace*{0.3cm}
Noting that the semi-circular density in Eq. (\ref{eq:free-density}) has a finite support over $x \in [-\sqrt{2N\sigma^2(\tau)}, + \sqrt{2N \sigma^2(\tau)}]$, where $\sigma^2(\tau) = D(1-e^{-2\mu \tau})/\mu$, the integral in (\ref{eq:def-FCS}) can be expressed in a compact form as
\bea \label{NL_1}
\frac{\langle N_L(\tau)\rangle}{N} \approx G\left( v = \frac{\ell}{\sqrt{1-e^{-2 \mu \tau}}}\right) \;,
\eea
where the scaling variable $v \geq \ell$ necessarily and the function $G(v)$ is given by (see also Fig. \ref{Fig_G} for a plot of this function)
\bea \label{def_G}
G(v) =
\begin{cases} 
&\frac{1}{\pi} v \sqrt{2-v^2} + \frac{2}{\pi} {\rm ArcTan}\left( \frac{v}{\sqrt{2-v^2}}\right) \quad, \quad \ell \leq v < \sqrt{2} \\
& \\
& 1 \quad, \quad \hspace*{5.1cm} v > \sqrt{2} \;.
\end{cases}
\eea
We have thus determined the first term in Eq. (\ref{FCS_3}) (which is ``deterministic'') explicitly for large $N$. The fluctuating second term (of order $O(\sqrt{\ln N})$) is much smaller compared to this ``deterministic'' term (of order $O(N)$). Hence to leading order for large $N$ we neglect the fluctuating term and express the distribution of $N_L(\tau)$ by
\bea \label{delta_NL}
{\rm Prob.}[N_L(\tau) = M ; \tau] \approx \delta\left( M - \langle N_L(\tau)\rangle \right) \;.
\eea
Substituting this in Eq. (\ref{FCS_2}) we get, for large $N$,
\bea \label{FCS_4}
{\rm Prob.}[N_L = M \vert r] \approx r\int_0^\infty \dd \tau  \,e^{-r \tau} \, \delta\left( M - \langle N_L(\tau)\rangle \right) \;.
\eea
We now substitute the explicit expression for $\langle N_L(\tau) \rangle$ in Eq. (\ref{NL_1}) into (\ref{FCS_4}) to write

\begin{figure}[t]
\centering
\includegraphics[width=0.32\textwidth]{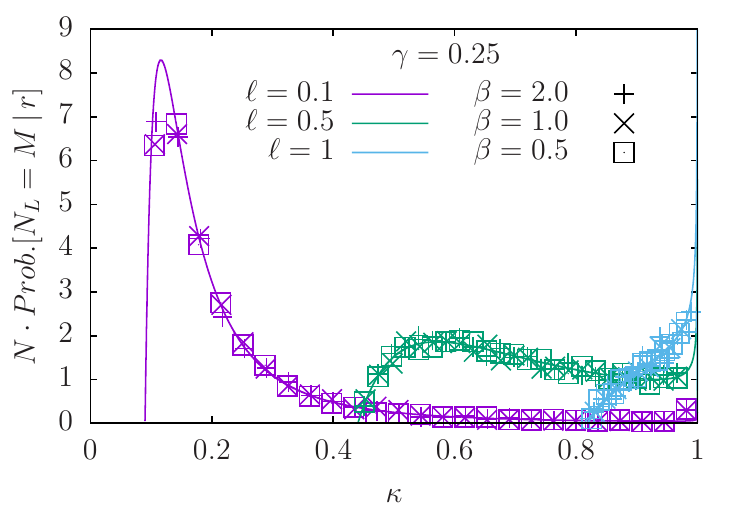}
\hfill
\includegraphics[width=0.32\textwidth]{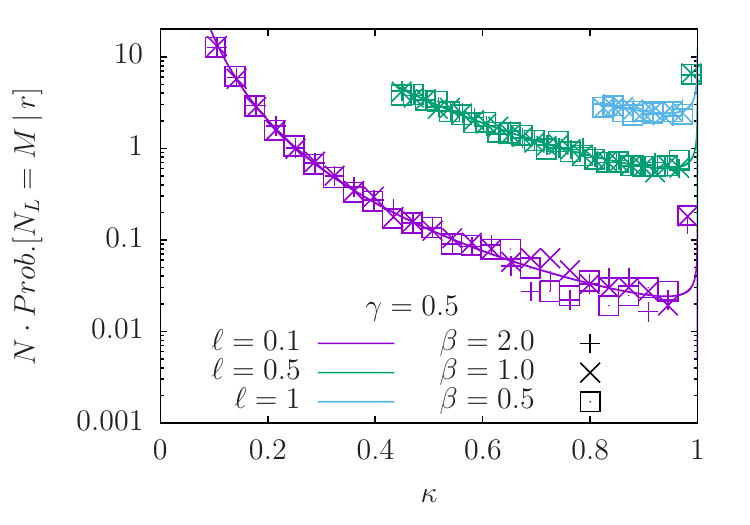}
\hfill
\includegraphics[width=0.32\textwidth]{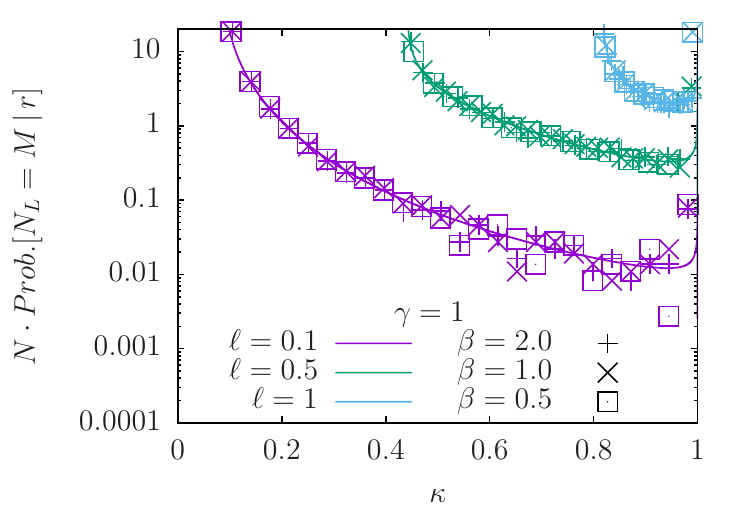}
\caption{Plot of the scaling function $q(\kappa, \ell, \gamma)$ vs $\kappa$ given in Eq. (\ref{eq:d-def-2}) describing the full counting statistics, i.e., the PDF of the number of particles in the interval $[-L, L]$. In this plot, we only show the smooth part of the scaling function in (\ref{eq:d-def-2}), i.e., without the delta function. 
The lines are the analytical predictions given by Eq. (\ref{eq:d-def-2}) while the points are obtained from numerical direct sampling simulations as described in the introduction of Section \ref{sec:model}. Different colors correspond to different values of $\ell$ (0.1, 0.5 and 1 for purple, green and blue) and different symbols correspond to different values of $\beta = 0.5,1,2$. The different panels correspond to different values of $\gamma = 1/4, 1/2, 1$ from left to right.}\label{fig:fcs}
\end{figure}
\begin{equation} \label{eq:2-interval-split}
{\rm Prob.}[N_L = M | r] = \frac{r}{N} \int_{I_1} \dd \tau \; e^{-r \tau} \delta\left(\frac{M}{N} - 1\right) + \frac{r}{N} \int_{I_2} \dd \tau \; e^{-r \tau} \delta\left(\frac{M}{N} - \frac{1}{N}\langle N_L(\tau) \rangle\right)  \;,
\end{equation}
where the integration domains are $I_1 = \left[0,-\ln(1-\ell^2/2)/(2 \mu)\right]$ and $I_2 = \left[-\ln(1-\ell^2/2)/(2 \mu),+\infty\right)$. Substituting Eq. (\ref{def_G}) we can express the distribution as 
\begin{equation} \label{eq:fcs-scaling}
{\rm Prob.}[N_L =M | r] \approx \frac{1}{N} \, q\left( \frac{M}{N}, L \sqrt{\frac{\mu}{N D}}, \mu/r \right) \;,
\end{equation}
where the scaling function $q(\kappa, \ell, \gamma)$ has support over $\kappa \in [\kappa_{\rm min}, 1]$ and is given by
\begin{align} 
q(\kappa, \ell, \gamma) = \delta(\kappa - 1) \left[1 - \left(1 - \frac{\ell^2}{2}\right)^{\frac{1}{2 \gamma}}\right] + \frac{\ell^2}{\gamma} \int_{\ell}^{\sqrt{2}} \frac{\dd v}{v^3} \left(1 - \frac{\ell^2}{v^2} \right)^{\frac{1}{2\gamma}-1} \delta\left( \kappa - G(v)\right)\;, \label{eq:d-def}
\end{align}
and
\begin{equation}\label{eq:def-zmin}
\kappa_{\rm min} = G(\ell) =\frac{1}{\pi} \ell \sqrt{2 - \ell^2} + \frac{2}{\pi} {\rm ArcTan}\left(\frac{\ell}{\sqrt{2 - \ell^2}}\right) \;.
\end{equation}
The lower limit $\kappa_{\rm min}$ comes from the limit $v \to \ell$ of the integrand in Eq. (\ref{eq:d-def}). To perform this integral, we look for the root $v_{\star} \equiv v_{\star}(\kappa)$ of the equation (see Fig. \ref{Fig_G})
\bea \label{eq:v-eq}
\kappa = G(v) \;,
\eea
in terms of which the scaling function $q(\kappa, \ell, \gamma)$ in Eq. (\ref{eq:d-def}) can be expressed as
\begin{align} \label{eq:d-def-2}
q(\kappa, \ell, \gamma) = \delta(\kappa - 1) \left[1 - \left(1 - \frac{\ell^2}{2}\right)^{\frac{1}{2 \gamma}}\right] + \frac{\pi \ell^2}{2 \sqrt{2} \, \gamma \, v_{\star}^3} \frac{1}{\sqrt{1-\frac{v_{\star}^2}{2}}} \left(1 - \frac{\ell^2}{v_{\star}^2} \right)^{\frac{1}{2 \gamma}-1} \;,
\end{align}
where we recall that $\ell < v_{\star}(\kappa) < \sqrt{2}$. 
%
%
%
%
%
%
Note that this scaling function $q(\kappa, \ell, \gamma)$ is normalized to unity, i.e., $\int_{\kappa_{\min}}^1 \dd \kappa \, q(\kappa, \ell, \gamma) = 1$ and is independent of $\beta$. One can easily work out the asymptotic behaviors of $q(\kappa, \ell, \gamma)$ as $\kappa$ approaches the two edges $\kappa = \kappa_{\min}$ and $\kappa = 1$. Skipping details, one can show that they are given by Eq. (\ref{eq:main-left-asymptotic-d}). We note in particular that the behavior of the scaling function 
$q(\kappa, \ell, \gamma)$ changes at $\gamma = 1/2$. For $\gamma < 1/2$ it vanishes as $\kappa \to \kappa_{\min}$, while it diverges for $\gamma > 1/2$.  
A plot of this scaling function is given in Fig. \ref{fig:fcs} and is in perfect agreement with numerical simulations.

\vspace*{0.3cm}
\noindent{\bf The variance of the number fluctuations.} We have computed the full distribution of $N_L$ above. However, a more experimentally accessible observable is the variance of $N_L$, characterising its fluctuations around the mean. Let us first recall how the variance behaves in the absence of resetting, i.e., for the standard log-gas with joint distribution given in Eq. (\ref{GOE}). In this case, the variance of $N_L$ was originally computed for $L = O(1/\sqrt{N})$, i.e., when $L$ is of the order of the inter-particle spacing \cite{DM63,CL95,FS95}. More recently, using Coulomb gas techniques, the variance has been computed for $L$ all the way up to the upper edge $L \sim O(\sqrt{N})$, for Gaussian random matrices and for arbitrary $\beta > 0$~\cite{MMSV14,MMSV16}. It has been shown that the variance ${\rm Var}(N_L) = \langle N_L^2\rangle - \langle N_L \rangle^2$ behaves as
\bea \label{var}
{\rm Var}(N_L) \approx \frac{1}{\beta \pi^2} \ln (N \ell (2-\ell^2)^{3/2}) \quad, \quad \ell < \sqrt{2} \;,
\eea   
where the dimensionless length $\ell$ is given by
\bea \label{ell}
\ell = L \sqrt{\frac{\mu}{N\,D}} \;.
\eea
Note that since the semi-circular density is supported over $[-\sqrt{2 N\,D/\mu}, +\sqrt{2 N\,D/\mu}]$, if the interval length $L$ exceeds $\sqrt{2 N\,D/\mu}$, i.e., $\ell > \sqrt{2}$, then the variance vanishes in the large $N$ limit, since all the particles are contained inside the support. Clearly the variance for large $N$ scales as $\ln N$, which is much smaller than the mean, which is of order~$O(N)$.  
\begin{figure}[t]
\includegraphics[width = 0.45 \linewidth]{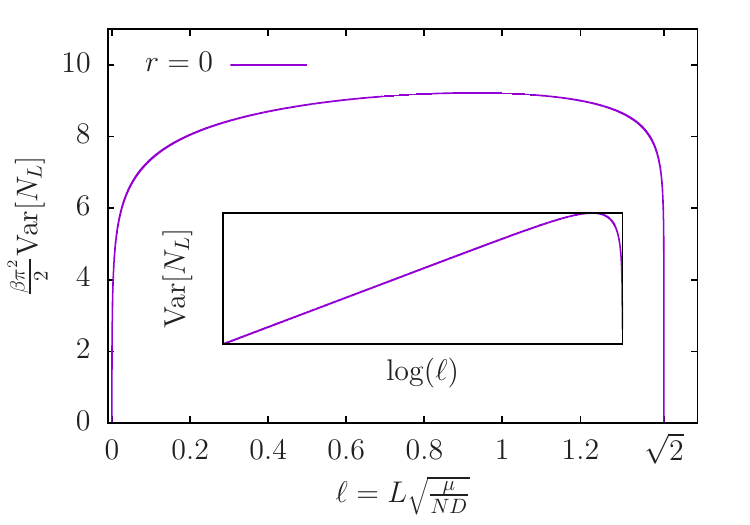}\includegraphics[width = 0.45 \linewidth]{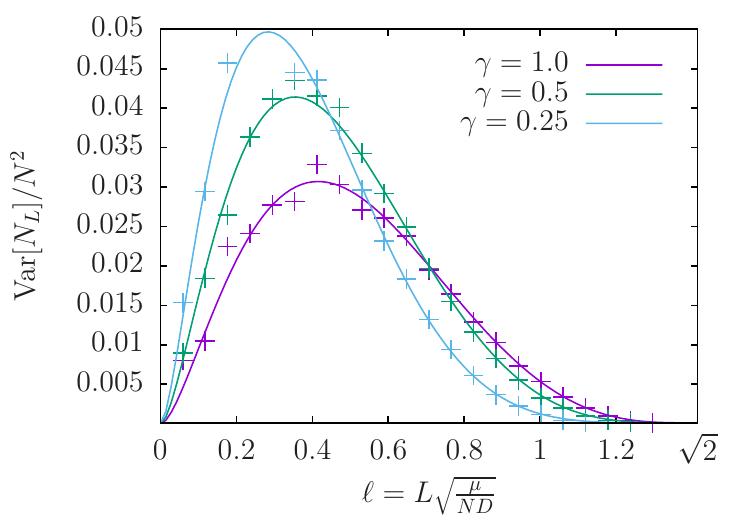}
\caption{On the left panel, we plot the variance of $N_L$ for $r=0$ as a function of the scaled distance $\ell$ given in Eq. (\ref{var}) for $N=1000$. The inset shows the same quantity but on semi-logarithmic scale. On the right panel, we plot the variance of $N_L$ for $r>0$ for three values of $\gamma$ as a function of the scaled distance $\ell$ as given in Eq. (\ref{Var_r_1}). The lines correspond to analytical predictions from Eq.~(\ref{Var_r_1}), while the symbols correspond to numerical simulations performed for $\beta = 2$.}\label{Fig:variance}
\end{figure}

\vspace*{0.3cm}
What happens when we switch on the resetting, i.e., for $r>0$? From Eq. (\ref{FCS_4}), we see that, to leading order for large $N$,
\bea \label{NLsq}
\langle N_L^2 \rangle \approx r\int_0^\infty \dd \tau e^{-r\tau}\, \langle N_L(\tau)\rangle^2  \;,
\eea
where $\langle N_L(\tau)\rangle$ is given in Eq. (\ref{NL_1}). Consequently, the variance is given by
\bea \label{Var_r}
{\rm Var}(N_L) = \langle N_L^2\rangle - \langle N_L \rangle^2 \approx  r\int_0^\infty \dd \tau e^{-r\tau}\, \left[\langle N_L(\tau)\rangle\right]^2 -  \left(r\int_0^\infty \dd \tau e^{-r\tau}\, \langle N_L(\tau)\rangle \right)^2 \;.
\eea 
Substituting the expression for $\langle N_L(\tau)\rangle$ from Eq. (\ref{NL_1}) and carrying out the integral we get, for large $N$,
\bea \label{Var_r_1}
{\rm Var}(N_L) \approx N^2 {\cal F}\left( \ell = L \sqrt{\frac{\mu}{N D}}, \gamma = \frac{\mu}{r}\right) \;,
\eea
where the scaling function ${\cal F}(\ell, \gamma)$, for fixed $\gamma$, is supported over $\ell \in [0,\sqrt{2}]$ and is independent of $\beta$. While it can be written explicitly, it has a rather long expression, which we do not provide here. 
However, for fixed $\gamma$, one can extract the asymptotic behaviors of $\mathcal{F}(\ell, \gamma)$ as $\ell$ approaches the edges of the support $[0,\sqrt{2}]$. One finds
\begin{equation}
\mathcal{F}(\ell, \gamma) \sim \begin{dcases}
B\, \left(\sqrt{2}-\ell \right)^{3+\frac{1}{2\gamma}} 
 &\mbox{~~when~~} \ell \to \sqrt{2} \;, \\
 & \\
- \frac{8 \ell^2 \log \ell }{\gamma \pi^2} &\mbox{~~when~~} \ell \to 0 \;,
\end{dcases}
\end{equation}
where $B = {2^{\frac{19}{2} + \frac{1}{4\gamma}}} \gamma^3/(3 \pi ^2(1+12 \gamma + 44 \gamma^2 + 48 \gamma^3))$. The scaling function $\mathcal{F}(\ell, \gamma)$ can be easily plotted (see the right panel of Fig. \ref{Fig:variance}). From Eq. (\ref{Var_r_1}), we see that, in the presence of resetting, the variance scales as $N^2$, as opposed to $\ln N$ in the absence of resetting. Thus fluctuations in $N_L$ become much more dominant for $r>0$, compared to the hyperuniform fluctuations for $r=0$. In Fig. \ref{Fig:variance}, we plot the variance as a function of the scaled distance $\ell$, both for $r=0$ (left panel) and $r>0$ (right panel). One sees very different behaviors in the two cases.

\begin{figure}[t]
\centering
\includegraphics[width=0.5\textwidth]{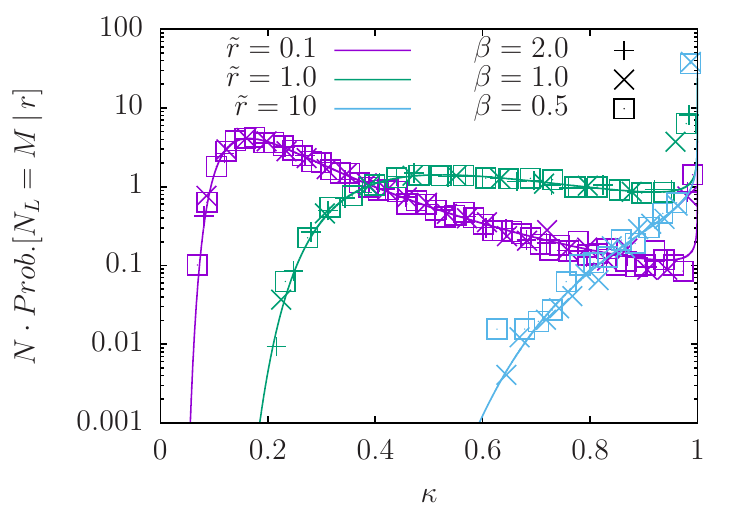}
\caption{Plot of the scaling function $q_0(\kappa, \tilde r)$ vs $\kappa = M/N$, for fixed $\tilde r = r L^2/(N\,D)$, given 
in Eq. (\ref{eq:free-q}) describing the PDF of the number of particles within a box $[-L, L]$ around the origin. The lines are the analytical predictions given by Eq. (\ref{eq:free-q}) -- only the smooth part, i.e., without the delta function -- while the symbols are obtained from numerical direct sampling simulations as described in the introduction of Section \ref{sec:model}. Different colors correspond to different values of $\tilde r$ (0.1, 1 and 10 for purple, green and blue) and different symbols correspond to different values of $\beta = 0.5,1,2$.}\label{fig:fcs-free}
\end{figure}

\vspace*{0.3cm}
\noindent{\bf The limit $\mu \to 0$.}
To obtain the limit $\mu = 0$, we have to take $\gamma \to 0$ and $\ell \to 0$ keeping $\ell^2 / \gamma = \tilde r = r L^2 / (N D)$ fixed. In this case, the distribution in Eq. (\ref{eq:fcs-scaling}) reduces to
\begin{equation} \label{eq:fcs-scaling_mu0}
{\rm Prob.}[N_L =M | r] \approx \frac{1}{N} \, q_0\left( \kappa=\frac{M}{N},\tilde r = \frac{rL^2}{N D} \right) \;,
\end{equation}
where the scaling function $q_0(\kappa, \tilde r)$, independent of $\beta$, is given by
\begin{eqnarray} \label{eq:free-q}
q_0(\kappa, \tilde r) &=& \delta(\kappa - 1) \left[ 1 - e^{-\tilde r/4} \right] + \tilde r \int_0^{\sqrt{2}} \frac{dv}{v^3} e^{-\frac{\tilde r}{2 v^2}}\, \delta\left(\kappa - G(v)\right) \nonumber \\
&=& \delta(\kappa - 1) \left[ 1 - e^{-\tilde r/4} \right]  + \frac{\pi}{2}\frac{\tilde r}{v_\star^3}   \frac{e^{-\frac{\tilde r}{2 v_\star^2}}}{\sqrt{2 - v_\star^2}} \;,
\end{eqnarray}
where we recall that $v_\star$ is the root of $G(v)=\kappa$ as in Eq. (\ref{eq:v-eq}).
%
One also sees from Eq. (\ref{eq:def-zmin}) that, as $\ell \to 0$, the lower edge $\kappa_{\min} \to 0$. This indicates that the scaling function in Eq. (\ref{eq:free-q}) is now supported on $\kappa \in [0, 1]$. The asymptotic behavior of $v_\star$ when $\kappa \to 0$ and $\kappa \to 1$ can be extracted from the expression of $G(v)$ in Eq. (\ref{def_G}) and they read
\bea \label{asympt_vstar}
v_\star \approx
\begin{cases}
&\dfrac{\pi}{2\sqrt{2}} \, \kappa \quad, \quad \quad \quad \hspace*{2.2cm}\kappa \to 0 \;,\\ 
& \\
& \sqrt{2} - \left(\dfrac{3 \pi}{2^{11/4}} \right)^{2/3}\, (1-\kappa)^{2/3} \quad, \quad \kappa \to 1 \;.
\end{cases}
\eea
Using these results, one can extract the asymptotic behaviors of $q_0(\kappa, \tilde r)$ 
in Eq. (\ref{eq:free-q}) as $\kappa \to 0$ and $\kappa \to 1$. They read
\bea \label{eq:free-left-asymp}
q_0(\kappa, \tilde r) \approx
\begin{cases}
& \dfrac{8 \tilde r}{\pi^2 \kappa^3} \exp[-\frac{4 \tilde r}{\pi^2 \kappa^2}] \quad, \quad \quad \quad \quad \hspace*{3.5cm}\; \kappa \to 0 \\
& \\
&(1-e^{-\tilde r/4})\,\delta(\kappa - 1) +  \dfrac{\pi^{2/3} \tilde r}{4 \times 6^{1/3}} e^{-\tilde r/4} (1-\kappa)^{-1/3} \quad, \quad \kappa \to 1 \;.
\end{cases}
\eea
We plot the scaling function $q_0(\kappa, \tilde r)$ in Fig.~\ref{fig:fcs-free} and compare it to numerical simulations, finding excellent agreement. Clearly, it has a very different shape compared to the $\mu > 0$ case. In particular, the asymptotic behaviors for $\mu=0$ in Eq. (\ref{eq:free-left-asymp}) can be compared with the corresponding asymptotics for $\mu >0$ in Eq. (\ref{eq:main-left-asymptotic-d}). One sees that, while the divergence as $(1-\kappa)^{-1/3}$ as $\kappa \to 1$ is similar for $\mu =0$ and $\mu >0$, the behaviours at the lower edge as $\kappa \to \kappa_{\min}$ are quite different in the two cases. It vanishes as a power law for $\mu > 0$ as $\kappa \to \kappa_{\min}$, while for $\mu=0$ it vanishes much more rapidly in an essential singular way as $\kappa \to 0$. 

\begin{figure}[t]
\includegraphics[width = 0.5 \linewidth]{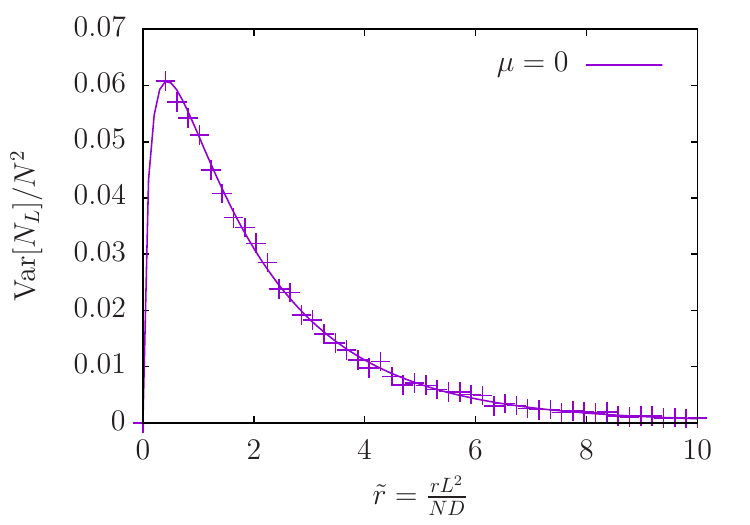}
\caption{Plot of the scaling function ${\cal F}_0(\tilde r)$ vs $\tilde r$, given in Eq. (\ref{var_mu0}), for the untrapped case $\mu = 0$. The line corresponds to the analytically obtained scaling function, while the symbols correspond to numerical simulations performed for $N = 1000$ and $\beta =2$.}\label{Fig:variancemu0}
\end{figure}

\vspace*{0.3cm}
Similarly, the variance of $N_L$ can also be extracted from Eq. (\ref{Var_r_1}) by taking the simultaneous limit $\gamma \to 0$, $\ell \to 0$ while keeping the  ratio $\ell^2/\gamma = \tilde r = r\,L^2/(N D)$ fixed. In this limit, one obtains
\bea \label{var_mu0}
{\rm Var}(N_L) \approx N^2 {\cal F}_0\left(\tilde r = \frac{r\, L^2}{N D}\right) \;,
\eea
where the scaling function ${\cal F}_0(\tilde r)$ still has a rather complicated form but its asymptotic behaviors can be computed exactly, leading to 
\begin{equation}
\mathcal{F}_0(\tilde r) \sim \begin{dcases}
 \frac{-4 \tilde r \log \tilde r}{\pi^2} &\mbox{~~when~~} \tilde r \to 0 \\
\frac{512}{3 \pi^2 \tilde r^3} e^{-\tilde r/4} &\mbox{~~when~~} \tilde r \to \infty \;.
\end{dcases}
\end{equation}
A plot of this scaling function is provided in Fig. \ref{Fig:variancemu0}.

\section{Conclusion} \label{sec:conclusion}

The $\beta$-DBM process refers to the stochastic evolution of a system of particles in a confined harmonic trap and with pairwise repulsive
interaction given in Eq. (\ref{Langevin}) and with a parameter $\beta>0$ characterising the relative strength of the interaction and the thermal noise. 
For $\beta = 1,2,4$, they can be related to the eigenvalues of a Gaussian random matrix whose entries perform independent Ornstein-Uhlenbeck processes.  
The joint distribution of the positions of the particles approaches at long times a stationary distribution that can be 
expressed in a Gibbs-Boltzmann form with an energy given in Eq.~(\ref{energy}). In this paper we have studied the $\beta$-DBM process for arbitrary $\beta$ but in the presence of stochastic resetting with a constant rate $r$, whereby
all the particles are simultaneously reset to $0$ and the process restarts immediately after. We call this new process {\it resetting Dyson Brownian motion} (RDBM) with parameter $\beta > 0$, i.e., the $\beta$-RDBM process. The stochastic resetting drives this $\beta$-RDBM process to a nonequilibrium stationary state (NESS) where, in addition to the repulsive interaction, the particles experience an all-to-all attraction generated dynamically by the simultaneous resetting. In this paper we have computed the joint distribution of the postions of the particles exactly in the NESS of the $\beta$-RDBM process and shown that it is rather different from the equilibrium measure of Dyson log gas. For instance, while on an average the particles are supported over a finite region of space in both models,
the fluctuations of the particle positions around this average are drastically different. The log-gas is known to be rather {\it rigid}, i.e., the fluctuations are small compared to the mean value for different observables, such as the extreme value statistics (EVS), i.e., the position of the rightmost particle. In contrast, we find that the $\beta$-RDBM in its NESS is rather {\it fluffy}, i.e., the fluctuations are of the same order as the average values. We have demonstrated this difference by calculating analytically (and verified numerically) several observables of the $\beta$-RDBM in its NESS, both macroscopic as well as microscopic. These include the average density profile of the gas, the EVS, the spacing distribution between two consecutive particles and the full counting statistics, i.e., the distribution of the number of particles in an interval $[-L,+L]$. 
 
\vspace*{0.3cm} 
Our results provide a new exactly solvable NESS of an interacting particle system subjected to stochastic resetting. These results also open up the possibilities for studying other models of interacting particles in the presence of stochastic resetting.  
For instance, one can also study the effect of stochastic resetting for DBM in the presence of geometric constraints such as a hard wall at the origin at $x=0$, which has been studied previously in the absence of resetting \cite{M1984,RS11,Bor09,NM09}. It would be interesting to study the NESS of such a system when driven by stochastic resetting. Another interesting system corresponds to having a type-dependent boundary, for which the reset-free DBM has been studied in Ref. \cite{Gautie}. How does such a system behave in the presence of resetting? 
Finally, in this paper, we have studied 
the limiting behaviour of the distribution of several observables in the NESS of the $\beta$-RDBM when $N \to \infty$. It would be interesting to compute the finite $N$ corrections, in particular the large deviation tails of these distributions.

\begin{acknowledgements}
We acknowledge support from ANR Grant No. ANR-23-CE30-0020-01 EDIPS.
\end{acknowledgements}

\newpage

\begin{appendix}

\section{Exact propagator of the $\beta$-DBM process at all times}\label{App}

\noindent The stochastic differential equation of the $\beta$-DBM process is given in Eq. (\ref{Langevin}) and it reads
\begin{equation} \label{eq:SDE-DBM}
\dv{x_i}{t} = -\mu x_i + D \sum_{j \neq i} \frac{1}{x_i - x_j} + \sqrt{\frac{2 D}{\beta}} \eta_i(t) \;,
\end{equation}
where $\eta_i(t)$ are delta-correlated independent Gaussian white noises with $\langle \eta_i(t) \rangle = 0$ and $\langle \eta_i(t)\eta_j(t') \rangle = \delta_{ij} \delta(t - t')$. For compactness we denote the joint probability distribution of finding the particles at $x_1, \cdots, x_N$ at time $t$ by
\begin{equation} \label{eq:def-propagator}
{\rm Prob.}[x_1, \cdots, x_N, t] = p(\vec{x}, t) \;.
\end{equation}
The Fokker-Planck equation of the process described in Eq. (\ref{eq:SDE-DBM}) is 
\begin{equation} \label{eq:FP-DBM}
\pdv{p(\vec{x}, t)}{t} = \frac{D}{\beta} \sum_{i = 1}^N \pdv[2]{p(\vec{x}, t)}{x_i} - \sum_{i = 1}^N \pdv{}{x_i} \left[\left(-\mu x_i + D \sum_{j \neq i} \frac{1}{x_i - x_j} \right) p(\vec{x}, t) \right] \;.
\end{equation}
At $t=0$ we assume that the particles start from the initial condition (\ref{IC}) in the main text, with $\epsilon \to 0$. 
As seen in the main text, for $\beta = 1, 2$ or $4$, the $\beta$-DBM process can be interpreted as the time evolution of the eigenvalues of an $N \times N$ Gaussian random matrix whose entries perform independent Ornstein-Uhlenbeck processes. This connection allows one to write the propagator at all times for $\beta = 1,2,4$ as \cite{F10} 
\begin{equation} \label{eq:ansatz-propagator}
p(\vec{x}, t) = \frac{1}{Z(\beta)} \frac{1}{\sigma(t)^{N + \beta N(N-1)/2}} \exp[ -\frac{\beta}{2}\left( \frac{1}{\sigma(t)^2} \sum_{i = 1}^N x_i^2 - \sum_{j \neq i} \log|x_i - x_j| \right) ] \;,
\end{equation}
where 
\begin{equation}
\sigma^2(t) = \frac{D}{\mu} \left(1 - e^{-2 \mu t}\right) \;.
\end{equation}

We now show that the same expression for the time-dependent propagator as in Eq. (\ref{eq:ansatz-propagator}) actually holds for all $\beta > 0$. 
To establish this, we simply substitute the expression in (\ref{eq:ansatz-propagator}) into the Fokker-Planck equation (\ref{eq:FP-DBM}) and verify that it does satisfy the Fokker-Planck equation at all times $t$. Upon this substitution, the left hand side (LHS) of Fokker-Planck equation (\ref{eq:FP-DBM}) reads
\begin{align}\label{LHS1}
{\rm LHS} = p(\vec{x}, t) \left[ \frac{\beta}{\sigma(t)^3} \sum_{i = 1}^N x_i^2 - \frac{N + \beta N(N-1)/2}{\sigma(t)} \right] & \left(\frac{D}{\sigma(t)} - \mu\sigma(t)\right) \;.
\end{align}
On the other hand, the right hand side (RHS) gives
\begin{align} \label{RHS1}
{\rm RHS} = p(\vec{x}, t) \sum_{i = 1}^N \Bigg\{ \frac{D}{\beta} &\left[ \left( -\frac{\beta}{\sigma(t)^2} x_i + \beta \sum_{j \neq i} \frac{1}{x_i - x_j} \right)^2 - \frac{\beta}{\sigma(t)^2} - \beta \sum_{j \neq i} \frac{1}{(x_i - x_j)^2}  \right] \nonumber\\
&+ \mu + D \sum_{j \neq i} \frac{1}{(x_i - x_j)^2} \nonumber\\
&- \left( -\mu x_i + D \sum_{j \neq i} \frac{1}{x_i - x_j} \right)\left(-\frac{\beta}{\sigma(t)^2} x_i + \beta \sum_{j\neq i} \frac{1}{x_i - x_j} \right) \Bigg\} \;.
\end{align}
Remarkably, the RHS simplifies further to
\begin{align} \label{RHS2}
{\rm RHS} =\left(\frac{D}{\sigma(t)^2} - \mu\right) \sum_{i = 1}^N \left\{ \frac{\beta}{\sigma(t)^2} x_i^2 -  \left(\beta \sum_{j \neq i} \frac{x_i}{x_i - x_j}  + 1 \right) \right\} 
\end{align}
We next use the following identity 
\begin{equation} \label{eq:sum-identity}
\sum_{i = 1}^N \sum_{j =1,j\neq i}^N \frac{x_i}{x_i - x_j} = \frac{N(N-1)}{2} \;,
\end{equation}
which can be proved simply as follows. Denoting this double sum by $\alpha$, we note that  
\bea \label{def_alpha}
\alpha = \sum_{i = 1}^N \sum_{j =1,j\neq i}^N \frac{x_i}{x_i - x_j} =  \sum_{i = 1}^N \sum_{j =1,j\neq i}^N \frac{x_i-x_j+x_j}{x_i - x_j} =  \sum_{i = 1}^N \sum_{j =1,j\neq i}^N \left(1 - \frac{x_i}{x_i-x_j} \right)\;,
\eea
where we interchanged the indices $i$ and $j$ in the last term. Noting that this last term is just $-\alpha$, it follows that $2 \alpha = N(N-1)$, leading to the identity in Eq. (\ref{eq:sum-identity}). Using this identity in Eq. (\ref{RHS2}) and comparing to the LHS in Eq. (\ref{LHS1}) it follows that LHS = RHS. 
Therefore the propagator given in Eq. (\ref{eq:ansatz-propagator}) satisfies the Fokker-Planck equation at all times $t$ and for all $\beta > 0$. 

\end{appendix}

\end{document}